\documentstyle[12pt]{article}

  \def\pp{{\mathchoice
          {
              \kern 1pt%
              \raise 1pt
              \vbox{\hrule width5pt height0.4pt depth0pt
                    \kern -2pt
                    \hbox{\kern 2.3pt
                          \vrule width0.4pt height6pt depth0pt
                          }
                    \kern -2pt
                    \hrule width5pt height0.4pt depth0pt}%
                    \kern 1pt
           }
            {
              \kern 1pt%
              \raise 1pt
              \vbox{\hrule width4.3pt height0.4pt depth0pt
                    \kern -1.8pt
                    \hbox{\kern 1.95pt
                          \vrule width0.4pt height5.4pt depth0pt
                          }
                    \kern -1.8pt
                    \hrule width4.3pt height0.4pt depth0pt}%
                    \kern 1pt
            }
            {
              \kern 0.5pt%
              \raise 1pt
              \vbox{\hrule width4.0pt height0.3pt depth0pt
                    \kern -1.9pt  
                    \hbox{\kern 1.85pt
                          \vrule width0.3pt height5.7pt depth0pt
                          }
                    \kern -1.9pt
                    \hrule width4.0pt height0.3pt depth0pt}%
                    \kern 0.5pt
            }
            {
              \kern 0.5pt%
              \raise 1pt
              \vbox{\hrule width3.6pt height0.3pt depth0pt
                    \kern -1.5pt
                    \hbox{\kern 1.65pt
                          \vrule width0.3pt height4.5pt depth0pt
                          }
                    \kern -1.5pt
                    \hrule width3.6pt height0.3pt depth0pt}%
                    \kern 0.5pt
            }
        }}

  \def\mm{{\mathchoice
                       {
                             \kern 1pt
               \raise 1pt    \vbox{\hrule width5pt height0.4pt depth0pt
                                  \kern 2pt
                                  \hrule width5pt height0.4pt depth0pt}
                             \kern 1pt}
                       {
                            \kern 1pt
               \raise 1pt \vbox{\hrule width4.3pt height0.4pt depth0pt
                                  \kern 1.8pt
                                  \hrule width4.3pt height0.4pt depth0pt}
                             \kern 1pt}
                       {
                            \kern 0.5pt
               \raise 1pt
                            \vbox{\hrule width4.0pt height0.3pt depth0pt
                                  \kern 1.9pt
                                  \hrule width4.0pt height0.3pt depth0pt}
                            \kern 1pt}
                       {
                           \kern 0.5pt
             \raise 1pt  \vbox{\hrule width3.6pt height0.3pt depth0pt
                                  \kern 1.5pt
                                  \hrule width3.6pt height0.3pt depth0pt}
                           \kern 0.5pt}
                       }}

\catcode`@=11
\def\un#1{\relax\ifmmode\@@underline#1\else
        $\@@underline{\hbox{#1}}$\relax\fi}
\catcode`@=12

\let\du=\du                     

\def\a{\alpha}
\def\b{\beta}
\def\c{\chi}
\def\d{\delta}

\def\f{\phi}
\def\g{\gamma}
\def\h{\eta}

\def\j{\psi}
\def\k{\kappa}
\def\l{\lambda}
\def\m{\mu}

\def\o{\omega}

\def\q{\theta}

\def\t{\tau}

\def\L{\Lambda}
\def\O{\Omega}

\def\ve{\varepsilon}
\def\vf{\varphi}

\def\bo{{\raise-.5ex\hbox{\large$\Box$}}}               
\def\pa{\partial}                                       
\def\de{\nabla}                                         
\def\pr{\prod}                                          
\def\TH{{\raise.2ex\hbox{$\displaystyle \bigodot$}\mskip-4.7mu \llap H \;}}
\def\face{{\raise.2ex\hbox{$\displaystyle \bigodot$}\mskip-2.2mu \llap {$\ddot
        \smile$}}}                                      

\def\leftrightarrowfill{$\mathsurround=0pt \mathord\leftarrow \mkern-6mu
        \cleaders\hbox{$\mkern-2mu \mathord- \mkern-2mu$}\hfill
        \mkern-6mu \mathord\rightarrow$}
\def\dvec#1{\vbox{\ialign{##\crcr
        \leftrightarrowfill\crcr\noalign{\kern-1pt\nointerlineskip}
        $\hfil\displaystyle{#1}\hfil$\crcr}}}           

\def\frac#1#2{{\textstyle{#1\over\vphantom2\smash{\raise.20ex
        \hbox{$\scriptstyle{#2}$}}}}}                   
\def\sfrac#1#2{{\vphantom1\smash{\lower.5ex\hbox{\small$#1$}}\over
        \vphantom1\smash{\raise.4ex\hbox{\small$#2$}}}} 
\def\bfrac#1#2{{\vphantom1\smash{\lower.5ex\hbox{$#1$}}\over
        \vphantom1\smash{\raise.3ex\hbox{$#2$}}}}       
\def\afrac#1#2{{\vphantom1\smash{\lower.5ex\hbox{$#1$}}\over#2}}    

\def\[{\lfloor{\hskip 0.35pt}\!\!\!\lceil}
\def\]{\rfloor{\hskip 0.35pt}\!\!\!\rceil}
\def\Lag{{\cal L}}
\def\du#1#2{_{#1}{}^{#2}}
\def\ud#1#2{^{#1}{}_{#2}}
\def\dud#1#2#3{_{#1}{}^{#2}{}_{#3}}
\def\udu#1#2#3{^{#1}{}_{#2}{}^{#3}}

\def\fracm#1#2{\hbox{\large{${\frac{{#1}}{{#2}}}$}}}

\def\tr{{\rm tr}}

\def\un{\underline}
\def\fracmm#1#2{{{#1}\over{#2}}}

\def\low#1{{\raise -3pt\hbox{${\hskip 0.75pt}\!_{#1}$}}}

\newskip\humongous \humongous=0pt plus 1000pt minus 1000pt
\def\caja{\mathsurround=0pt}
\def\eqalign#1{\,\vcenter{\openup2\jot \caja
        \ialign{\strut \hfil$\displaystyle{##}$&$
        \displaystyle{{}##}$\hfil\crcr#1\crcr}}\,}
\newif\ifdtup

\def\pl#1#2#3{Phys.~Lett.~{\bf {#1}B} (19{#2}) #3}
\def\np#1#2#3{Nucl.~Phys.~{\bf B{#1}} (19{#2}) #3}

\def\pr#1#2#3{Phys.~Rev.~{\bf D{#1}} (19{#2}) #3}
\def\cqg#1#2#3{Class.~and Quantum Grav.~{\bf {#1}} (19{#2}) #3}

\def\mpl#1#2#3{Mod.~Phys.~Lett.~{\bf A{#1}} (19{#2}) #3}

\def\ibid#1#2#3{{\it ibid.}~{\bf {#1}} (19{#2}) #3}

\parskip=\medskipamount                 
\thicklines                         

\begin{document}


\thispagestyle{empty}               

\def\border{                                            
        \setlength{\unitlength}{1mm}
        \newcount\xco
        \newcount\yco
        \xco=-24
        \yco=12
        \begin{picture}(140,0)
        \put(-20,11){\tiny Institut f\"ur Theoretische Physik Universit\"at
Hannover~~ Institut f\"ur Theoretische Physik Universit\"at Hannover~~
Institut f\"ur Theoretische Physik Hannover}
        \put(-20,-241.5){\tiny Institut f\"ur Theoretische Physik Universit\"at
Hannover~~ Institut f\"ur Theoretische Physik Universit\"at Hannover~~
Institut f\"ur Theoretische Physik Hannover}
        \end{picture}
        \par\vskip-8mm}

\def\headpic{                                           
        \indent
        \setlength{\unitlength}{.8mm}
        \thinlines
        \par
        \begin{picture}(9,16)
        \put(75,16){\line(1,0){4}}
        \put(80,16){\line(1,0){4}}
        \put(85,16){\line(1,0){4}}
        \put(92,16){\line(1,0){4}}

        \put(85,0){\line(1,0){4}}
        \put(89,8){\line(1,0){3}}
        \put(92,0){\line(1,0){4}}

        \put(85,0){\line(0,1){16}}
        \put(96,0){\line(0,1){16}}
        \put(92,16){\line(1,0){4}}

        \put(85,0){\line(1,0){4}}
        \put(89,8){\line(1,0){3}}
        \put(92,0){\line(1,0){4}}

        \put(85,0){\line(0,1){16}}
        \put(96,0){\line(0,1){16}}
        \put(79,0){\line(0,1){16}}
        \put(80,0){\line(0,1){16}}
        \put(89,0){\line(0,1){16}}
        \put(92,0){\line(0,1){16}}
        \put(79,16){\oval(8,32)[bl]}
        \put(80,16){\oval(8,32)[br]}

        \end{picture}
        \par\vskip-6.5mm
        \thicklines}

\border\headpic {\hbox to\hsize{
\vbox{\noindent DESY 97 -- 247  \hfill December 1997 \\
ITP--UH--35/97 \hfill hep-th/9712151 }}}

\noindent
\vskip1.3cm
\begin{center}

{\Large\bf An action of $N=8$ self-dual supergravity  
\vglue.1in
            in ultra-hyperbolic harmonic superspace
\footnote{Supported in part by the `Deutsche Forschungsgemeinschaft' and the
NATO grant CQG 930789}}\\
\vglue.3in

Sini\v{s}a Karnas and Sergei V. Ketov~\footnote{
On leave of absence from:
High Current Electronics Institute of the Russian Academy of Sciences,
\newline ${~~~~~}$ Siberian Branch, Akademichesky~4, Tomsk 634055, Russia}

{\it Institut f\"ur Theoretische Physik, Universit\"at Hannover}\\
{\it Appelstra\ss{}e 2, 30167 Hannover, Germany}\\
{\sl ketov@itp.uni-hannover.de}
\end{center}
\vglue.2in
\begin{center}
{\Large\bf Abstract}
\end{center}

\noindent
The $N$-extended self-dual supergravity in the ultra-hyperbolic four-dimensional
spacetime of kleinian signature (2+2) is given in the $N$-extended harmonic 
superspace. We reformulate the on-shell $N$-extended self-dual supergravity 
constraints of Siegel to a `zero-curvature' representation, and solve all of 
them but one in terms of a single superfield prepotential, by using a covariant 
Frobenius gauge in the Devchand-Ogievetsky approach. An off-shell superspace action, 
whose equation of motion yields
the remaining constraint, is found. Our manifestly Lorentz-covariant action in 
harmonic superspace is very similar to the non-covariant Chern-Simons-type 
action, which was proposed earlier by Siegel in the light-cone $N=8$ superspace.
Our action is also invariant under the residual superdiffeomorphisms and the 
residual local $OSp(8|2)$ super-Lorentz rotations, which are left after 
imposing the Frobenius gauge. The infinitesimal superfield parameters of the 
residual symmetries are expressed in terms of independent analytic superfields.

\newpage

\section{Introduction}

The {\it self-dual Yang-Mills} (SDYM) and {\it self-dual gravity} (SDG) 
in four euclidean 
spacetime dimensions are usually associated with the instanton solutions to 
the corresponding non-self-dual classical field theories, which result in a 
finite euclidean action. The SDYM field strengh and the SDG curvature tensor,
$$ F={}^*F~,\quad {\rm and}\quad  R={}^*R~,\eqno(1.1)$$
imply the equations of motion for a gauge field and metric, respectively, just
because of the Bianchi identities. Hence, the notion of self-duality is 
stronger than that of the equations of motion, and one may wonder whether an 
off-shell action exists which would yield the self-duality equations, with all 
the linear symmetries to be manifestly realised in the action. The 
reason why we would like to find covariant actions for self-dual field theories
is closely related to the problem of an off-shell covariant description of 
BPS-like field configurations, or branes generalizing the instantons in various 
dimensions.

An educated answer to the problem of constructing the actions for self-dual field
theories is crucially dependent upon answering the following two related
questions first, namely,
\begin{itemize}
\item should the action in question have the same symmetries as a given   
self-duality condition~?
\item should the number of physical degrees of freedom in an off-shell action 
be kept, or it is allowed to increase them~?
\end{itemize}
In other words, on the one hand, there may exist many off-shell actions, which
give {\it in particular} the desired self-duality relation in some 
{\it gauge-fixed} form, as it is usually the case. On the other hand, naive
approaches, e.g. when using Lagrange multipliers, easily do the job but they
may simultaneously lead to some additional propagating degrees of freedom whose
decoupling may require extra constraints, or result in a trivial (free) theory
in the case of self-duality. Therefore, if one insists on maintaining all the 
linearly realised symmetries {\it and\/} the number of degrees of freedom, the 
problem of formulating a Lorentz-covariant action for a given self-duality 
condition becomes non-trivial, and it is not obvious whether a solution exists
at all.  

Supersymmetry adds new interesting aspects to this problem~\cite{sie,kgn}.
The four-dimensional self-duality turns out to be closely related to the 
existence of Majorana-Weyl (i.e. real chiral) spinors, which only exist in
the ultra-hyperbolic spacetime of kleinian signature (2+2). 
The SDYM and SDG have essentially one state of `helicity' $(-1)$ and $(-2)$, 
respectively.~\footnote{In the 
SDYM case, this state is Lie algebra valued.} When being treated off-shell, 
the Lorentz invariance already implies a need for another state with the 
opposite `helicity' to compensate the otherwise negative `helicity' of an 
off-shell action. Of course, it looks like adding a Lagrange multiplier, 
while such field is already present in the case of the {\it maximally} extended 
$N=4$ {\it self-dual super-Yang-Mills} (SDSYM) and the $N=8$ 
{\it self-dual supergravity} (SDSG). These are just the only two self-conjugate 
cases where all particles come in pairs, so that each field in the action can 
naturally serve as a Lagrange multiplier for another field with the opposite 
`helicity'. 

The manifestly Lorentz-covariant action for the $N=4$ SDSYM theory in 
components was given by Siegel~\cite{sie}. He also found the manifestly $N=4$
supersymmetric (but non-covariant) superfield action for the same theory in
light-cone superspace~\cite{sie}. An off-shell $N=4$ SDSYM action, which would
be both manifeslty Lorentz-covariant and $N=4$ supersymmetric, seems to exist 
in the harmonic superspace {\it only}, i.e. with an infinite number of 
auxiliary fields. Such action was found by Sokatchev in ref.~\cite{eme}, where
it takes the form of the Chern-Simons-type action in terms of properly chosen 
gauge connections in $N=4$ harmonic superspace. It is the main purpose of this
paper to generalise the result of ref.~\cite{eme} to the case of the $N=8$ 
SDSG. 

The $(N=0)$ SDG in the ultra-hyperbolic spacetime of kleinian signature (2+2) is 
also known as the effective field theory of interacting $(2,2)$ {\it spinning}
 strings, i.e. the critical closed strings with the $(2,2)$ 
world-sheet supersymmetry~\cite{ov}. Because of a `topological' nature of the 
spinning strings, their non-vanishing amplitudes seem to be only 3-point 
functions~\cite{bv}. This very basic observation already imples the cubic 
(i.e. Chern-Simons) -type of the effective self-interaction (see subsect.~2.1 
for more). The effective action of the $(2,2)$ spinning strings is, however, 
not fully Lorentz-invariant with respect to the $SO(2,2)$ rotations, just like
the spinning string amplitudes themselves. It is a particular, gauge-fixed and
non-covariant (Pleba\~nski) version of SDG that actually appears to be the 
effective field theory of the $(2,2)$ closed spinning strings~\cite{ov}. If 
the effective theory were fully $SO(2,2)$ covariant, this would imply an 
equivalence between the $(4,4)$ and $(2,2)$ spinning strings, as well as a
maximally supersymmetric spectrum of particles in the target four-dimensional 
ultra-hyperbolic `spacetime'. However, since the $(4,4)$ and $(2,2)$ spinning 
strings have different critical dimensions, and there are no extra fermionic 
particles interacting with a single physical state of `helicity' $(-2)$ in the 
spectrum of $(2,2)$ spinning strings~\cite{bkl}, no full Lorentz invariance and 
no any supersymmetry are allowed in the four-dimensional target (`spacetime') of
the critical closed $(2,2)$ strings. Even if one takes into account a possible
inequivalence of different pictures or superconformal ghost vacua which, 
in principle, may lead to an infinite number of massless physical states in the 
spinning string theory~\cite{inf}, all these states cannot be spacetime 
fermionic because of the spectral flow associated with two-dimensional N=2 
superconformal algebras~\cite{ss}. Nevertheless, the effective full Lorentz 
invariance, as well as the effective $(4,4)$ world-sheet (twisted) 
supersymmetry, do appear in the topological reformulation of the interacting 
closed $(2,2)$ spinning string theory in the twistor space, not in the
`spacetime'~\cite{bv}. 

Because of the existence of $N$-extended supergravitites~\cite{sie,kgn}, 
one may ask about a possible existence
of yet another non-trivial `spinning' or `heterotic' closed string theory, 
which would lead to an $N$-extended SDSG as its effective field theory in the
string target space.~\footnote{See e.g., ref.~\cite{ber} for some recent 
efforts in this direction, as regards the $N=1$ SDSG.} To our knowledge, no
string theory is known, which would have the $N=8$ SDSG as its effective field
theory. The covariant component action similar to that of the $N=8$ SDSG may 
arise as a consistent finite truncation of the most general effective action of 
the critical closed spinning strings with the extra massless physical states 
corresponding to the inequivalent pictures as in ref.~\cite{inf}, but with the
bosonic states only and no `spacetime' supersymmetry in the conventional 
(Lie superalgebra) sense. The alternative may be a possible relevance of the 
maximally extended $N=8$ SDSG to the F-theory branes~\cite{ket}. 

It should also be mentioned that the four-dimensional SDYM and SDG equations
of motion may also be considered as the master integrable systems because of
the fact that they provide a natural classification scheme for many integrable
systems in lower dimensions~\cite{mw}. Along these lines, the SDSYM and SDSG
can be understood as yet another extensions of integrable systems, whose 
possible significance for the non-perturbative string theories 
(= M- and F-theories) is yet to be understood.

Our paper is organized as follows: in sect.~2 we briefly discuss SDG and its 
relation to the $(2,2)$ critical closed strings, formulate the standard
on-shell constraints defining the $N=8$ gauged SDSG in superspace, and 
reformulate the constraints to harmonic superspace. Sect.~3 is devoted to a 
comparison of the $N=8$ SDSG in harmonic superspace with the $N=4$ SDSYM 
in the formulation of Sokatchev~\cite{eme}. In sects.~4 and 5 we collect some 
technical details about the Devchand-Ogievetsky approach to self-duality 
\cite{doe,do}, and define a Frobenius gauge. The prepotentials in the
Frobenius gauge are also introduced in  sect.~5, while the action is formulated in 
sect.~6. We summarize our conclusions in Sect.~7. For the sake of completeness,
the known component results about the $N=8$ SDSG are collected in Appendix.
\vglue.2in

\section{From SDG to its supersymmetric $N$-extended (SDSG) generalization}

Since the $(N=0)$ SDG in the four-dimensional ultra-hyperbolic spacetime, whose
signature is (2+2) and the natural Lorentz group is $SO(2,2)$, is going to be
our starting point, we would like to remind the reader some basic facts about
SDG and its relation to the spinning strings in subsect.~2.1. The on-shell 
formulation of the $N=8$ SDSG \cite{sie} is given in subsect.~2.2. A 
reformulation of the on-shell $N=8$ SDSG in harmonic superspace is discussed in
subsect.~2.3 along the lines of ref.~\cite{do}. The whole sect.~2 serves as a
technical introduction for the next sections.

\subsection{SDG and closed spinning strings}

Because of the isomorphism 
$$ SO(2,2)\cong SL(2,{\bf R})\otimes SL(2,{\bf R})'~,\eqno(2.1)$$
it is natural to represent four ultra-hyperbolic spacetime coordinates as
$x^{\a\a'}$, where the spinor indices $\a=(1,2)$ and $\a'=(1',2')$ refer to 
$SL(2)$ and $SL(2)'$, respectively.

A commutator of the (curved) spacetime covariant derivatives defines the 
curvature tensor~\cite{do},
$$ \[ \de_{\b\b'},\de_{\a\a'}\]=\ve_{\a'\b'}R_{\a\b}+\ve_{\a\b}R_{\a'\b'}~,
\eqno(2.2)$$
where $R_{\a\b}$ can be decomposed with respect to the generators $(M^{\g\d},
M^{\g'\d'})$ of $SL(2)\otimes SL(2)'$ as follows:
$$ R_{\a\b}=C_{(\a\b\g\d)}M^{\g\d} +R_{\a\b(\a'\b')}M^{\a'\b'}+\fracm{1}{6}
RM_{\a\b}~,\eqno(2.3)$$
and similarly for $R_{\a'\b'}$, 
$$ R_{\a'\b'}=C_{(\a'\b'\g'\d')}M^{\g'\d'} +R_{\g\d(\a'\b')}M^{\g\d}
+\fracm{1}{6} RM_{\a'\b'}~.\eqno(2.4)$$
Here $C_{(\a'\b'\g'\d')}$ $(C_{(\a\b\g\d)})$ are the totally symmetric 
components of the (anti)-self-dual Weyl tensor, $R_{(\a\b)(\a'\b')}$ are the
components of the traceless Ricci tensor, and $R$ is the scalar 
curvature.~\footnote{Any symmetrization of indices (in brackets) is defined
with unit weight.} In this notation, the self-duality of the Riemann curvature
in eq.~(1.1) just means
$$ R_{\a'\b'}=0~,\eqno(2.5)$$
or, equivalently, because of eq.~(2.4),
$$ C_{(\a'\b'\g'\d')}=0~,\eqno(2.6)$$
and
$$ R_{\g\d(\a'\b')}=0~,\quad {\rm and}\quad R=0~.\eqno(2.7)$$
Eq.~(2.7) is equivalent to the Einstein equations without matter, whereas the
vanishing of the self-dual Weyl tensor in eq.~(2.6) represents the only 
additional condition needed for self-duality.

In the case of SDG, eqs.~(2.2) and (2.3) simplify to
$$  \[ \de_{\b\b'},\de_{\a\a'}\]=\ve_{\a'\b'}R_{\a\b} \eqno(2.8)$$
and
$$ R_{\a\b}=C_{(\a\b\g\d)}M^{\g\d}~,\eqno(2.9)$$
respectively. It is now obvious that the self-dual curvature takes its values
in the $sl(2)$ algebra only. In other words, it is the $SL(2)$ factor of the 
full Lorentz group $SO(2,2)$ that should be promoted to the local symmetry,
whereas the $SL(2)'$ symmetry should be kept global or rigid because of 
eq.~(2.5). Therefore, we consider the first spinor index $(\m)$ of the 
ultra-hyperbolic spacetime coordinate $x^{\m\a'}$ as a curved (world) index,
and keep another spinor index $\a'$ as a flat (tangential) index associated
with the flat ($sl(2)'$-valued) part of the curvature. It leads to the 
following form of the covariant derivative
$$ \de_{\a\a'}=E^{\m\b'}_{\a\a'}\pa_{\m\b'} +\o_{\a\a'}~,\eqno(2.10)$$
where a `vierbein' $E^{\m\b'}_{\a\a'}$ and a connection
$$\o_{\a\a'}=\o_{\a\a'\b\g}M^{\b\g}\eqno(2.11)$$
have been introduced. Since the self-duality implies no restrictions on the
$sl(2)$-valued curvature $R_{\a\b}$, the solutions to SDG amount to the 
vierbeins which have no torsion. In other words, it is the vanishing
torsion constraint that is the SDG equation of motion in this approach. 

As is well-known, the self-duality in four dimensions is equivalent to a
Ricci-flat K\"ahler geometry~\cite{ahit}.
 A K\"ahler geometry has a complex structure,
while keeping the complex structure intact is only compatible with a part
(subgroup) $U(1,1)$ of the full Lorentz group transformations $SO(2,2)\cong
SU(1,1)\otimes SU(1,1)$. Since the NSR formulation of the spinning string 
theory with the $(2,2)$ world-sheet supersymmetry is intrinsically complex, 
i.e. it actually requires
choosing a complex structure for its definition, it is not very surprising that
the spinning string theory is only invaraint under the subgroup $U(1,1)$ of
$SO(2,2)$, even in the case of a flat background spacetime~\cite{ov,bkl}. The
missing symmetry $SO(2,2)/U(1,1)~\sim~SU(1,1)/GL(1,{\bf R})$ can be understood
 as the
twistor transformations rotating the complex structure. It is now clear why
the full Lorentz symmetry can be formally restored in the extended twistor
space~\cite{bv} but not in the ultra-hyperbolic spacetime which is the target
space of a spinning string.

Given a complex structure $x^{\a\a'}\to (z^i,\bar{z}^{\bar{i}})$, $i=1,2$, and
a K\"ahler potential of the form
$$ K=K_0+4\k^3\f, \quad{\rm where}\quad 
K_0=\h_{i\bar{j}}z^i\bar{z}^{\bar{j}}~, \quad 
\h_{i\bar{j}}=\h^{i\bar{j}}=\left(\begin{array}{cc}
+1 & 0 \\ 0 & -1\end{array}\right)~,\eqno(2.12)$$
and $\k$ is the gravitational coupling constant of the inverse mass dimension,
the Ricci-flatness condition takes the form
$$ \det(g_{i\bar{j}})=-1~,\eqno(2.13)$$
in terms of the K\"ahler metric
$$ g_{i\bar{j}}=\pa_i\pa_{\bar{j}}K=\h_{i\bar{j}}
+4\k^3\pa_i\pa_{\bar{j}}\f~,\eqno(2.14)$$
where we have used the identity for the Ricci tensor in a K\"ahler geometry,
$R_{i\bar{j}}=\pa_i\pa_{\bar{j}}\log\det (g_{k\bar{l}})$ and the boundary
condition $g_{i\bar{j}}\to\h_{i\bar{j}}$ at infinity. In terms of the K\"ahler
deformation $\f$, eq.~(2.13) can be obtained as an equation of motion from the
following (Pleba\~nski) action
$$S_{\rm Plebanski}=\int d^{2+2}z\,\left(\fracmm{1}{2} \h^{i\bar{j}}\pa_i\f
\bar{\pa}_{\bar{j}}\f +\fracmm{2\k^3}{3}\f\pa\bar{\pa}\f\wedge\pa\bar{\pa}\f
\right)~.\eqno(2.15)$$
As was shown by Ooguri and Vafa~\cite{ov}, it is exactly the action (2.15) that
also appears to be the effective field theory action of the closed spinning 
strings. The Pleba\~nski action (2.15) is only invariant under the $U(1,1)$ 
part of the $SO(2,2)$ Lorentz transformations. Moreover, this action has a 
dimensionful coupling constant $\k$ which is absent in the Lorentz-covaraint
equation (1.1) for the SDG.

\subsection{On-shell SDSG in superspace}

The $N$-extended supersymmetrization  of SDG essentially amounts to extending
the $SL(2)$ local symmetry to the $OSp(N|2)$ local symmetry,~\footnote{The
superconformal symmetry would imply $SO(2,2)\stackrel{conf}{\longrightarrow}
SO(3,3)\cong SL(4)\stackrel{susy}{\longrightarrow}SL(N|4)$.} 
while keeping the manifest global symmetry $SL(2)'$ intact. 
Since the supergroup $OSp(N|2)$ contains the $SO(N)$
Lie group in addition, this leads to the {\it gauged} version of the 
$N$-extended SDSG, in which the internal symmetry $SO(N)$ rotating $N$
supercharges is gauged. The {\it on-shell} formulations of the $N$-extended
SDSG in superspace are similar for all $N$.

Let $g$ be the corresponding (dimensionless) gauge coupling constant (the
gravitational coupling constant $\k$ of the inverse mass dimension was already
introduced in the previous subsect.~2.1). Let $\h^{ab}$, $a,b=1,2,\ldots,N\,$, 
be the $so(N)$ Cartan-Weyl metric multiplied by the factor $g/\k$, i.e. of 
mass dimension, and $C^{\a\b}\sim\ve^{\a\b}$ the self-dual part of the charge 
conjugation matrix for spacetime spinors. The $OSp(N|2)$ metric can now be 
introduced as
$\h^{AB}=(\h^{ab},C^{\a\b})$ where $A=(a,\a)$. The Grassmann grading is 
defined by treating the $SL(2)$ and $SL(2)'$ spinor indices $\a$ and $\a'$
as bosonic indices, and the $SO(N)$ vector indices $a,b,\ldots$ as fermionic 
indices. The $OSp(N|2)$ generators $M^{AB}$ act on vectors $V^C$ as follows:
$$ \[ M^{AB},C^C \} = V^{[A}\h^{B\}C}~,\eqno(2.16)$$
so that one has the identity~\cite{sie}
$$ \fracm{1}{2} K_{BC}\[M^{CB},V_A\}=K_{AB}\h^{CB}V_C \eqno(2.17)$$
for any matrix $K_{AB}$.

The on-shell constraints defining the $N$-extended SDSG in the $N$-extended
chiral  superspace $z^{M\a'}=(x^{\m\a'},\q^{m\a'})$ can be naturally divided
into two groups. The first group of (anti)commutators of the superspace 
covariant derivatives $\de^{a\a}$ and $\de_{b\b'}$ reads~\cite{sie}
$$ \{ \de^{a\a},\de^{b\b}\}=C^{\a\b}M^{ab} +\h^{ab}M^{\a\b}~,\eqno(2.18)$$
$$ \{\de^{a\a},\de_{b\b'}\}=\d^a_bC^{\a\b}\de_{\b\b'}~,\quad
\[ \de^{a\a},\de_{\b\b'}\]=\d^{\a}_{\b}\h^{ab}\de_{b\b'}~,$$
where $M^{ab}$ are the $SO(N)$ generators of dimension of mass. Eqs.~(2.18) 
naturally define the gauged $N$-extended local supersymmetry, while $M^{ab}$, 
$M^{\a\b}$ and $\de^{a\a}$ can be recognized as the generators $M^{AB}$ of 
$OSp(N|2)$. In particular, it implies that the spacetime covariant derivatives 
$\de^{a\a}$ should be interpreted as some of the $OSp(N|2)$ generators
\cite{sie}. The second group of the one-shell superspace constraints defining 
the $N$-extended SDSG is a rather straightforward generalization of the SDG 
equation (2.8) to the supersymmetric case, namely,
 $$ \[ \de_{A\a'},\de_{B\b'}\} =C_{\b'\a'}R_{AB}~, \eqno(2.19)$$
in terms of the supercovariant superspace derivatives ({\it cf.} eq.~(2.10))
$$ \de_{A\a'}=E^{M\b'}_{A\a'}\pa_{M\b'}+\fracm{1}{2}\O_{A\a' BC}M^{CB}~,
\eqno(2.20) $$
where a supervielbein $E^{M\b'}_{A\a'}$ and a superconnection $\O_{A\a' BC}$
have been introduced. More explicitly, eq.~(2.19) can be written down as 
follows:
$$\eqalign{
\{ \de_{a\a'},\de_{b\b'}\}~=& C_{\b'\a'}\f_{ab}~,\cr
\[ \de_{a\a'},\de_{\b\b'}\]~=& C_{\b'\a'}\c_{a\b}~,\cr
\[ \de_{\a\a'},\de_{\b\b'} \]~=& C_{\b'\a'}R_{\a\b}~,\cr}
\eqno(2.21)$$
where $(\f_{ab},\c_{a\b},R_{\a\b})=R_{AB}$ is the $OSp(N|2)$-valued 
supercurvature tensor.

The superspace formulation of an $N$-extended SDSG given above is manifestly 
$N$-supersymmetric, and it is invariant under the local $OSp(N|2)$ symmetry
by construction. It is, however, an on-shell formulation since eq.~(2.19) has 
no torsion on its right-hand-side which implies the equations of motion, as we
already know from the previous subsection. One of the ways to go off-shell is 
to turn to a light-cone superspace formulation of the same theory, where only 
physical (i.e. propagating) field components are kept. As was shown by 
Siegel~\cite{sie}, all the on-shell SDSG superspace constraints but one can be
solved in the light-cone formalism, in terms of a {\it single} $N$-extended 
superfield prepotential $V_{='='}$ of `helicity' $(-2)$. 
The $(N=0)$ SDG prepotential originates from the equation 
$$ C_{\a\b\g\d}(x)\sim\pa_{\a+'}\pa_{\b+'}\pa_{\g+'}\pa_{\d+'}V_{='='}(x)~,
\eqno(2.22)$$
and it can be identified (up to a constant dimensionful factor) with the 
`scalar' field $\f$ introduced in the previous subsect.~2.1. It is obvious 
how to generalize eq.~(2.22) in the self-dual $N$-extended superspace to
$$  C_{ABCD}(z)\sim\pa_{A+'}\pa_{B+'}\pa_{C+'}\pa_{D+'}V_{='='}(z)~.
\eqno(2.23)$$
The free field equation for the SDSG prepotential $V_{='='}(z)$~\cite{sie},
$$\pa\du{A}{\a'}\pa_{B\a'}V_{='='}=0~, \eqno(2.24)$$ 
can actually be solved for all the $\q^{a-'}$-dependent components. It implies
that $V_{=',='}(z)$ can be reduced in the light-cone formalism to a 
{\it self-dual} superfield $V_{='='}(x^{\a\a'},\q^{a+'})$ that merely 
depends upon a half (Majorana-Weyl) of the anticommuting superspace 
coordinates. As a result, all 
the constraints in the light-cone formalism can be reduced to a single 
non-covariant equation for the self-dual superfield prepotential $V_{='='}$, 
which can be obtained from the $N$-extended Pleba\~nski action
\cite{sie}
$$ S_{\rm l.-c.}=\int d^{2+2}x d^N \q\, \left[ \frac{1}{2}V_{='='}\bo
V_{='='} +\frac{i}{6}V_{='='}(\pa\ud{\a}{+'}\pa_{A+'}V_{='='})\h^{BA}
(\pa_{B+'}\pa_{\a+'}V_{='='})\right]~.\eqno(2.25)$$
This light-cone $N$-extended superspace action is manifestly supersymmetric
with respect to a half of the original on-shell supersymmetry since it is 
written down in terms of a half (Majorana-Weyl) of the anticommuting 
superspace coordinates. Though the $N=8$ action (2.25) is neutral with respect 
to the parabolic subgroup $GL(1,{\bf R})'$ of the  $SL(2,{\bf R})'$ Lorentz 
symmetry, it is still not covariant with respect to the full $SL(2,{\bf R})'$ 
symmetry which is explicitly broken in the light-cone approach. Though a 
covariant off-shell description of the $N=8$ SDSG exists in components 
(see ref.~\cite{sie} and our Appendix), it is not manifestly supersymmetric. Our
main purpose in this paper is to `covariantize' the light-cone $N=8$ SDSG action
(2.25) in harmonic superspace. 

\subsection{SDSG in harmonic superspace}

Since the $SL(2)'$ Lorentz symmetry remains a global symmetry in the on-shell
superspace formulation of the $N$-extended SDSG, while it is broken to a 
parabolic subgroup $GL(1)'$
in the off-shell light-cone formulation, it is natural to `covariantize' the 
light-cone theory by introducing extra (twistor) harmonic variables 
$u^{\a'\pm}$ valued in the coset $SL(2)'/GL(1)'$, i.e. 
$$ u^{\a'\pm}\in SL(2,{\bf R})' \quad {\rm and}\quad u^{\a'+}u_{\a'}^-=1~,
\eqno(2.26)$$
and then apply the formal rules of the harmonic superspace approach along the
lines of ref.~\cite{gikos} for this non-compact case. This procedure was
successfully applied for a construction of a covariant action of the $N=4$
SDSYM in harmonic superspace in ref.~\cite{eme}, and we are now going to
proceed in the similar way, in the case of the $N=8$ SDSG.

Harmonic functions with definite $GL(1)'$ charge $q\geq 0$ are formally defined
by their expansion in terms of the harmonic variables, i.e.
$$ F^{(q)}(u)=\sum_{n=0}^{\infty} f^{(\a'_1\cdots \a'_{n+q}\b'_1\cdots \b'_n)}
u^+_{\a'_1}\cdots u^+_{\a'_{n+q}}u^-_{\b'_1}\cdots u^-_{\b'_n}~,\eqno(2.27)$$
where $f^{\a'_1\cdots \b'_n}$ are $SL(2)'$ tensors of `spin' $n+\frac{1}{2}q$.
The harmonic covariant derivatives take the form
$$ \pa^{++}=u^{\a'+}\fracmm{\pa}{\pa u^{\a'-}}~,\quad
 \pa^{--}=u^{\a'-}\fracmm{\pa}{\pa u^{\a'+}}~,\quad
 \pa^{0}=u^{\a'+}\fracmm{\pa}{\pa u^{\a'+}}
-u^{\a'-}\fracmm{\pa}{\pa u^{\a'-}}~,\eqno(2.28)$$
and satisfy an $sl(2)$ algebra. It is not difficult to check that
$$\pa^{++}F^{(q)}(u)=0\quad{\rm implies}\quad \left\{ 
\begin{array}{lr} F^{(q)}(u)=0~, & {\rm when}~ q<0~,\\
F^{(q)}(u)=constant, & {\rm when}~ q=0~,\\
F^{(q)}(u)=f^{(\a'_1\cdots \a'_q)}u^+_{\a'_1}\cdots u^+_{\a'_q}~,  
& {\rm when}~ q>0~.
\end{array}\right. \eqno(2.29)$$
The integration rules are defined as follows~\cite{gikos}:
$$ \int du=1~,\quad \int du\, u^+_{(\a'_1}\cdots u^+_{\a'_n}u^-_{\b'_1}\cdots
u^-_{\b'_n)}=0~,\eqno(2.30)$$
so that they project out the singlet part of an integrand with vanishing 
$GL(1)'$ charge. An integration by parts is allowed since
$$ \int du\,\pa^{++}F^{--}(u)=0~.\eqno(2.31)$$

The availability of harmonic variables allows one to define the 
Lorentz-covariant $GL(1)'$ projections like $x^{\m\pm}=u^{\pm}_{\a'}x^{\m\a'}$
and $\q^{m\pm}=u^{\pm}_{\a'}\q^{m\a'}$, and similarly for the superspace
covariant derivatives. 

The on-shell $N$-extended SDSG was defined in subsect.~2.2 by the constraints 
(2.19) in terms of the curved superspace covariant derivatives (2.20) and the
structure group $OSp(N|2)$. By using the harmonic coordinates $u^{\a'\pm}$, 
the covariant derivatives can be rewritten to the form
$$ \de_A^{\pm}=u^{\a'\pm}\de_{A\a'}= e_A^{\pm}+\o_A^{\pm}~,\eqno(2.32)$$
where the (super)vielbeine
$$\eqalign{ e^+_A=& E^M_A\pa^+_M + E_A^{++M}\pa^-_M~,\cr
e^-_A=& F^M_A\pa^-_M + F_A^{--M}\pa^+_M~,\cr}\eqno(2.33)$$
have been introduced. The superconnections $\o_A^{\pm}$ are defined by
eq.~(2.32).

In the original ({\it central}) basis of the harmonic superspace, in terms of 
the central coordinates $z^{M\pm}=z^{M\a'}u^{\pm}_{\a'}$ and $u^{\pm}_{\a'}$, 
the superdiffeomorphisms are realised via the transformations
$$ \d z^{M\pm}=\l^{M\pm}(z^{P\pm},u)~,\quad \d u^{\pm}_{\a'}=0~,\eqno(2.34)$$
Hence, the harmonic covariant derivatives $(\de^{++},\de^{--},\de^0)$ in the 
central basis are still of the form (2.28), as in the flat harmonic superspace.

It is now straightforward to verify that the $N$-extended SDSG constraints can
be put into the form~\cite{do}
$$ \[ \de^+_A,\de^+_B \]  =0~,\eqno(2.35)$$
$$ \[ \de^{++},\de^+_A \] =0~,\eqno(2.36)$$
$$ \[ \de^+_A, \de^-_B \}  =R_{AB}~,\eqno(2.37)$$
$$ \[ \de^{++},\de^-_A \} =\de^+_A~.\eqno(2.38)$$
One gets eq.~(2.35) after contracting eq.~(2.19) with harmonics. In order to
get eq.~(2.19) back from eq.~(2.35), one needs eq.~(2.36) which implies that
the derivative $\de^+_A$ in the central coordinates is linear in the harmonics
$u^+$. The most general solution to eq.~(2.35) might, however, have torsion 
terms on the right-hand-side of eq.~(2.19). It is eq.~(2.37) that takes care 
of it, since the torsion terms would then also appear on the right-hand-side
of eq.~(2.37) too. The last eq.~(2.38) is just needed to make sure that the
covariant derivative $\de^-_A$ in the central coordinates is linear in $u^-$. 

To solve the extended superspace constraints, it is usually useful to make a
transform to the so-called {\it analytic} basis in harmonic superspace, which 
allows one to realise the relevant symmetries in a smaller {\it analytic} 
subspace~\cite{gikos}. It simultaneously implies introducing more gauge fields,
since the harmonic covariant derivatives in the analytic coordinates will no
longer be of the form (2.28).

The transform from the central coordinates to the analytic ones is usually
described in terms of a `bridge' function $b(z,u)$~\cite{gikos},
$$ z^{M\pm}_{\rm a}=z^{M\pm} + b^{M\pm}(z,u)~.\eqno(2.39)$$
The analytic subspace $(z^{M+}_{\rm a},u)$ is supposed to be invariant under 
the analytic superdiffeomorphisms, i.e.
$$ \d z^{M+}_{\rm a}=\l^{M+}(z^+_{\rm a},u)~,\eqno(2.40)$$
whereas 
$$\d z^{M-}_{\rm a}=\l^{M-}(z^+_{\rm a},z^-_{\rm a},u)~,\eqno(2.41)$$ 
in general. Eq.~(2.39) also implies
$$ \d b^{M\pm}=\d z_{\rm a}^{M\pm}-\d z^{M\pm}\equiv \l^{M\pm}-\t^{M\pm}~.
\eqno(2.42)$$

All the SDSG constraints (2.35)--(2.38) keep their form after the transform 
(2.39), since they were written down in terms of the most general covariant 
derivatives in a curved superspace. However, in the analytic coordinates, the 
harmonic covariant derivatives receive some extra terms, $\pa^{++}\to D^{++}$,
i.e.
$$ \eqalign{
D^{++} &=~ \pa^{++}(z_{\rm a}^{\pm},u)\fracmm{\pa}{\pa(z_{\rm a}^{\pm},u)} \cr
&=~\pa^{++} + H^{+M}\fracmm{\pa}{\pa z^{M-}_{\rm a}}+
H^{+3M}\fracmm{\pa}{\pa z^{M+}_{\rm a}} \cr
&=~ \pa^{++} + H^{+M}\pa^+_{{\rm a}\,M}+ H^{+3M}\pa^-_{{\rm a}\,M}~,\cr}
\eqno(2.43)$$
where we have introduced the harmonic vielbeine
$$\eqalign{
H^{+M}=&~ \pa^{++}(z^{-M}+b^{-M}) \cr
=&~ z^{+M}+\pa^{++}b^{-M} \cr
=&~ z_{\rm a}^{+M}-b^{+M}+\pa^{++}b^{-M} \cr}\eqno(2.44)$$
and
$$ \eqalign{
H^{+3M}=& \pa^{++}(z^{+M}+b^{+M}) \cr
=&~ \pa^{++}b^{+M}~,\cr}\eqno(2.45)$$
as well as the notation $\pa^+_{{\rm a}M}=\pa/\pa z^{-M}_{\rm a}$
and $\pa^-_{{\rm a}M}=\pa/\pa z^{+M}_{\rm a}$. The harmonics themselves remain
inert under superdiffeomorphisms. Similarly, the vielbein (2.33) gets 
transformed as
$$ \eqalign{
e^+_A &~\to~ (e^+_Az_{\rm a}^{M-})\pa^+_{{\rm a}M}+
(e^+_Az_{\rm a}^{M+})\pa^-_{{\rm a}M} \cr
&~\to~ (e^+_Az_{\rm a}^{M-})\pa^+_{{\rm a}M}~\to~ 
E^M_A\pa^+_{{\rm a}M}~,\cr}\eqno(2.46)$$
where we have used the fundamental property 
$$ e^+_Az^{+M}_{\rm a}=0 \eqno(2.47)$$
of the analytic coordinates. One finds similarly that
$$ \eqalign{
e^-_A &~\to~ (e^-_Az_{\rm a}^{M+})\pa^-_{{\rm a}M}+
(e^-_Az_{\rm a}^{M+})\pa^+_{{\rm a}M} \cr
&~\to~ F^M_A\pa^-_{{\rm a}M}+ F_{A}^{--M}\pa^+_{{\rm a}M}~.\cr}
\eqno(2.48)$$

It follows from the constraint (2.35) that there exists an $OSp(N|2)$ valued
superfield $\vf$ which satisfies
$$  \de^+_A\vf=(e^+_A+\o^+_A)\vf=0~.\eqno(2.49)$$
Hence, a solution to eq.~(2.35) takes the form
$$ \de^+_A=e^+_A-(e^+_A\vf)\vf^{-1}~,\eqno(2.50)$$
whose connection is trivial. Therefore, we can get 
$$\de^+_A~\to~ \vf^{-1}\de^+_A\vf= e^+_A \eqno(2.51)$$ 
via an $OSp(N|2)$ rotation $e^+_A\to e^+_A=(\f e^+)_A$. The matrix $\f$ thus
plays the role of a `bridge' between the $OSp(N|2)$ transformations in the 
central coordinates and that in the analytic ones. In particular, one finds 
for the harmonic covariant derivative that 
$$ \de^{++}~\to~\vf^{-1}\de^{++}\vf=D^{++}+\o^{++}~,\eqno(2.52)$$
where
$$ \o^{++}=\vf^{-1}D^{++}\vf~.\eqno(2.53)$$
Since the general $OSp(N|2)$ transformation law of the connection $\o^+_A$ is 
given by
$$ \d\o^+_A=D^+_A\L -\[\L,\o^+_A\]~,\eqno(2.54)$$
fixing the gauge $\o^+_A=0$ implies 
$$D^+_A\L=0~,\eqno(2.55)$$
i.e. the $OSp(N|2)$-valued transformation parameter $\L$ should an
analytic superfield. The transformation laws of all the vielbein superfields 
can be written down as follows:
$$ \eqalign{
\d E^M_A=&~ E^N_A\pa^+_N\l^{M-}+\l_A^B E^M_B~,\cr
\d F^M_A=&~ F^N_A\pa^-_N\l^{M+}+\l_A^B F^M_B~,\cr 
\d F_A^{--M}=&~ F^N_A\pa^-_N\l^{M-}+F_A^{--N}\pa^+_N\l^{M-}+
\l_A^B F_B^{--M}~,\cr
\d H^{+3M}=&~ D^{++}\l^{+M}~,\cr
\d H^{+M}=&~ D^{++}\l^{-M}~,\cr}
\eqno(2.56)$$
where the infinitesimal parameters  $\l^{M\pm}$ have been introduced in 
eqs.~(2.40) and (2.41), and
$\l^A_B$ are the infinitesimal parameters of local $OSp(N|2)$ rotations.

We emphasize that an analyticity condition like that in eq.~(2.55) is only 
covariant with respect to the local analytic transformations. In the next 
sect.~3 the SDSG constraints will be analyzed in the analytic representation.
Since in the rest of our paper we are going to deal with the analytic 
superspace coordinates only, we omit the subscript `a' in what follows. 

\section{$N=8$ ~SDSG versus ~$N=4$~ SDSYM}

The maximally extended $N=8$ SDSG and $N=4$ SDSYM are both self-conjugate in
the sense that all their physical states with opposite `helicities' can be 
naturally paired to form scalars. This implies the existence of a manifestly
covariant and supersymmetric action for these theories. Such action for the 
$N=4$ SDSYM was constructed by Sokatchev~\cite{eme}, by using partial gauges
to solve some of the SDSYM constraints in harmonic superspace and then find
an action for the rest of the constraints to be interpreted as the equations 
of motion. In subsect.~3.1  we recapitulate some of the results of 
ref.~\cite{eme}~\footnote{See also ref.~\cite{do}.} since our $N=8$ SDSG 
construction, in fact, follows the Sokatchev pattern for the $N=4$ SDSYM up to
a point where the differences between the SDSYM and the SDSG become important. 

\subsection{$N=4$~ SDSYM action in harmonic superspace}

The standard (on-shell) constraints defining the $N=4$ SYM theory in the flat
$N=4$ superspace $(x^{\a\a'},\q^{\a}_a,\q^{a\a'})$, where 
$a,b,\ldots=1,2,3,4$ are the indices of the $SL(4)$ automorphism group of 
$N=4$ supersymmetry, read in terms of the gauge-covariant and (flat) 
supercovariant derivatives as follows~\cite{eme}:
$$\eqalign{
\[ \de^a_{\a},\de^b_{\b}\]=\ve_{\a\b}\tilde{\f}^{ab}~, \quad &
\[ \de_{a\a'},\de_{b\b'}\]=\ve_{\a'\b'}\f_{ab}~, \cr
\{ \de^a_{\a},\de_{\b\b'}\}=\ve_{\a\b}\tilde{\c}^{a}_{\b'}~,\quad &
\{ \de_{a\a'},\de_{\b\b'}\}=\ve_{\a'\b'}\c_{a\b}~, \cr
\[\de^a_{\a},\de_{b\b'}\]=\d^a_b\de_{\a\b'}~,\quad &
\{ \de_{\a\a'},\de_{\b\b'} \}=\ve_{\a'\b'}F_{\a\b}+\ve_{\a\b}F_{\a'\b'}~,\cr}
\eqno(3.1)$$
where the real scalars $\f_{ab}$ and $\tilde{\f}^{ab}$ are related as
$$ \tilde{\f}^{ab}=\frac{1}{2}\ve^{abcd}\f_{cd}~.\eqno(3.2)$$
 
In the case of the $N=4$ SDSYM, a half of the superfield strengths on the 
right-hand-side of eq.~(3.1) vanishes, while there is no constraint (3.2). The
on-shell $N=4$ SDSYM constraints can be divided into two groups, namely,
$$ \[ \de^a_{\a},\de^b_{\b}\]=0~,\quad \[\de^a_{\a},\de_{b\b'}\]=\d^a_b
\de_{\a\b'}~,\quad \{\de^a_{\a},\de_{\b\b'}\}=0~,\eqno(3.3)$$
and
$$ \[ \de_{a\a'},\de_{b\b'}\]=\ve_{\a'\b'}\f_{ab}~,\quad
\{ \de_{a\a'},\de_{\b\b'}\}=\ve_{\a'\b'}\c_{a\b}~,\quad
\{ \de_{\a\a'},\de_{\b\b'}\}=\ve_{\a'\b'}F_{\a\b}~,\eqno(3.4)$$
which are similar to the SDSG eqs.~(2.18) and (2.21), respectively.

Harmonic superspace is useful in rewriting the $N=4$ SDSYM constraints above 
into a `zero-curvature' form which is easier to deal with. In particular, the 
SDSYM analogues to the SDSG harmonic projections in eq.~(2.32) are
$$  \de^+_a=u^{\a'+}\de_{a\a'}~,\quad {\rm and}\quad
\de^+_{\a}=u^{\a'+}\de_{\a\a'}~.\eqno(3.5)$$

It is quite natural to assume that both the gauge superfield parameters and
the superconnections of the SDSYM are dependent upon the harmonic variables, 
while their actual linear dependence in accordance to eq.~(3.5) can be 
recovered as a solution due to some additional constraints like that in the 
SDSG eqs.~(2.36) and (2.38). One therefore needs a trivial connection to 
the harmonic covaraint derivative $\pa^{++}$, i.e.   
$$ \de^{++}=\pa^{++}+A^{++}~,\eqno(3.6)$$
and the extra constraints
$$ \{\de^{++},\de^a_{\a}\}=0~,\quad \{\de^{++},\de^+_a\}=0~,\quad 
\{\de^{++},\de^+_{\a}\}=0~.\eqno(3.7)$$
The initial $N=4$ SDSYM constraints (3.3) and (3.4) can now be put into the 
following equivalent `zero-curvature' form~\cite{eme}:
$$ \[ \de^a_{\a},\de^b_{\b}\]=0~,\eqno(3.8)$$
$$ \[ \de^a_{\a},\de^+_{b}\]=\d^a_b\de^+_{\a}~,\eqno(3.9)$$
$$ \{ \de^a_{\a},\de^+_{\b}\}=0~,\eqno(3.10)$$
$$ \[ \de^+_{a},\de^+_{b}\]=0~,\eqno(3.11)$$
$$ \[ \de^+_{\a},\de^+_{\b}\]=0~,\eqno(3.12)$$
$$ \[ \de^+_{\a},\de^+_{\b}\]=0~.\eqno(3.13)$$

The central idea of ref.~\cite{eme} was the use of a certain supersymmetric 
gauge, which is a combination of a chiral and a semi-analytic gauges.  The 
chiral gauge 
$$ \de^a_{\a}=\pa^a_{\a}~,\quad {\rm or} \quad A^a_{\a}=0~,\eqno(3.14)$$
is allowed due to eq.~(3.8). Together with eq.~(3.10), it implies
the chirality of the gauge superfields $A^+_{\a}$ and $A^{++}$, i.e. their
independence upon $\q^{\a}_a$. Eq.~(3.9) can now be solved as
$$ A^+_a=a^+_a(x,\q^{a\pm},u) +\q^{\a}_a A^+_{\a}(x,\q^{a\pm},u) \eqno(3.15)$$
in terms of two chiral superfields $a^+_a$ and $A^+_{\a}$. After substituting 
eq.~(3.15) into eq.~(3.11) one finds
$$\eqalign{
\pa^+_a a^+_b +\pa^+_ba^+_a +\{a^+_a,a^+_b\}&=~0~,\cr
\pa^+_a A^+_{\b} -\pa^+_{\b}a^+_a +\[a^+_a,A^+_{\b}\]&=~0~,\cr
\pa^+_{\a} A^+_{\b} -\pa^+_{\b}A^+_{\a} +\[A^+_{\a},A^+_{\b}\]&=~0~.\cr}
\eqno(3.16)$$
The first line of eq.~(3.16) allows one to impose a supersymmetric 
semi-analytic gauge
$$ a^+_a=0~,\eqno(3.17)$$  
in addition to the chiral gauge (3.14). The rest of equations (3.16), in fact,
allows one to fully gauge away $A^+_a$, i.e. take $A^+_{\a}=0$ also. This 
twistor transform is useful in discussing the solutions to the $N=4$ SDSYM 
equations~\cite{do}, but it turns out to be too restrictive for writing 
down an off-shell action~\cite{eme}.

In the chiral semi-analytic gauge, the second eqs.~(3.7) and (3.16) imply 
that the remaining harmonic gauge superfields $A^+_{\a}$ and $A^{++}$ are 
chiral and analytic simultaneously, i.e. they are only dependent upon 
$(x,\q^{a+},u)$, whereas the remaining constraints
$$\eqalign{
\pa^+_{\a}A^{++} -D^{++}A^+_{\a}+\[A^+_{\a},A^{++}\]&=~0~,\cr
\pa^+_{\a} A^+_{\b} -\pa^+_{\b}A^+_{\a} +\[A^+_{\a},A^+_{\b}\]&=~0~,\cr}
\eqno(3.19)$$
appear as the equations of motion for them. Eqs.~(3.19) are obviously 
invariant under the gauge transformations
$$ \eqalign{
\d A^+_{\a}&=~\pa^+_{\a}\L +\[A^+_{\a},\L\]~,\cr
\d A^{++}&=~D^{++}\L +\[A^{++},\L\]~,\cr}
\eqno(3.20)$$
whose gauge parameters $\L(x,\q^{a+},u)$ are chiral analytic superfields too.

The action, whose variational equations with respect to the 
independent superfields $A^+_{\a}$ and $A^{++}$ give the $N=4$ SDSYM equations
(3.20), is given by~\cite{eme}
$$ S_{N=4~SDSYM}= \int d^{2+2}xd^4\q^+du\,\tr\left(
A^{++}\pa^{\a+}A^+_{\a}-\frac{1}{2}A^{\a+}D^{++}A^+_{\a}+A^{++}A^{\a+}A^+_{\a}
\right)~.\eqno(3.21)$$
This Chern-Simons-type action of the $N=4$ SDSYM theory is fully covariant and 
$N=4$ supersymmetric, and it has no dimensionful coupling constant. In the rest 
of our paper, our main goal will be to formulate a similar action for the $N=8$ 
SDSG. 

\subsection{SDSG in a chirally analytic harmonic superspace}

Some of the vanishing SDSG (anti)commutators are quite similar to that of the 
SDSYM. For instance, eq.~(2.35) is equivalent to
$$ \{ \nabla^{+}_{a},\nabla^{+}_{b}\}=0~,\quad
\[ \nabla^{+}_{a},\nabla^{+}_{\beta}\]=0~,\quad
\[ \nabla^{+}_{\alpha},\nabla^{+}_{\beta}\]=0~, \eqno(3.22)$$
whereas eq.~(2.36) amounts to
$$ \[ \nabla^{++},\nabla^{+}_{a}\]=0~,\quad
\[ \nabla^{++},\nabla^{+}_{\alpha}\]=0~.\eqno(3.23)$$

The first eq.~(3.22) can be solved as in the previous subsect.~3.1, i.e. 
by imposing a gauge $\de^+_a=\pa^+_a$, where the full covariant derivative 
$\de^+_a$ is given by
$$ \de^+_a=e^+_a +\o^+_a~,\eqno(3.24)$$
in terms of the vielbein $e^+_a$ and the connection $\o^+_a$. Under the
local $OSp(N|2)$ rotations this covariant derivative transforms as
$$\eqalign{
\delta\nabla^{+}_{a}~=&\[ e^{+}_{a}+\omega^{+}_{a},\Lambda \] \cr
~=&\[ e^{+}_{a},\lambda_{CD}M^{DC}\] + \[ \omega^{+}_{a},\Lambda\] \cr
~=&e^{+}_{a}(\lambda_{CD}M^{DC})-\lambda_{CD}(M^{DC}e^{+}_{a})-\[ \Lambda,
\omega^{+}_{a} \] \cr
~=&(e^{+}_{a}\lambda_{CD})M^{DC}+(-)^{a(C+D)}\lambda_{CD}(e^{+}_{a}M^{DC})
-\lambda\_{CD}(M^{DC}e^{+}_{a})-\[ \Lambda,\omega^{+}_{a} \] \cr
~\equiv & D^{+}_{a}\Lambda-\lambda_{CD}\[ M^{DC},e^{+}_{a}\] - \[ \Lambda,
\omega^{+}_{a} \] \cr
~=&D^{+}_{a}\Lambda-2\lambda_{aC}\eta^{DC}e^{+}_{D}+\[ \omega^{+}_{a},\Lambda
 \]~, \cr}$$
where we have introduced the `short' derivative $D^+_a$ (without a connection) 
for our convenience. The connection $\omega^+_a$ transforms as usual,
$$ \delta\omega^{+}_{a}=D^{+}_{a}\Lambda+\[\omega^{+}_{a},\Lambda\]~.\eqno(3.25)$$
However, it follows from this equation that the gauge fixing $\omega^{+}_{a}=0$ 
implies a restriction $D^{+}_{a}\Lambda=0$. This means that the infinitesimal 
parameter $\L$ of the super-Lorentz rotations should be a chirally analytic 
superfield in the gauge $\de^+_a=\pa^+_a$, similarly to the supergauge parameter
of the SDSYM (subsect.~3.1). 

It should be noticed that our gauge for the vierbein, $e^+_a=\d^m_a\pa^+_m$ or
$E^m_a=\d^m_a$, does not imply the vanishing curvature in SDSG. However, since 
the vierbein is not inert under the $OSp(N|2)$ gauge transformations and the 
general coordinate transformations, 
$$ \delta E_{a}^{m}=\delta\delta_{a}^{m}=E^n_a(\pa^+_n\l^{-m})+
\lambda_{aC}\eta^{DC}E^{m}_{D}=\pa^+_a\l^{-m}+
\lambda_{ac}\eta^{dc}\d_{d}^m +\lambda_{a\alpha}
C^{\beta\alpha}E^{m}_{\beta}~, \eqno(3.26)$$
our gauge fixing also implies that the infinitesimal parameters $\l^{-m}$ of the 
general coordinate transformations and $\l_{ab}$ of the super-Lorentz 
transformations are to be related, whereas the parameters $\l_{a\a}$ of the 
local supersymmetry vanish, $\l_{a\a}=0$.

It is now obvious that the edition of a chirally analytic harmonic superspace, 
in the form used for the SDSYM, has to be $OSp(N|2)$ covariantized in the case of
SDSG. The main reason is the supersymmetric nature of the super-Lorentz structure
`group' $OSp(N|2)$ which is, in fact, a supergroup that mixes bosonic and 
fermionic tangent space indices. This simultaneously tells us what should be done
in the case of the SDSG, namely, the bosonic and fermionic tangent space indices 
are to be democratically treated within a single superindex. In other words, we 
should impose a super-Lorentz covariant gauge $e^+_A=\pa^+_A$ instead of the
non-covariant one used above, and relax our analyticity conditions. As will
be shown in the next sections, this proposal allows us to formulate an action for
the $N=8$ SDSG. 

\section{SDSG constraints and Frobenius gauge}

In this section we calculate the (anti)commutators of the $N$-extended SDSG in
some detail. This is needed in order to reduce the number of superfields, as well
as the number of symmetries acting on the superfields. We also introduce a
Frobenius gauge~\cite{do}, which will allow us to introduce SDSG superfield 
prepotentials in the next sect.~5. We follow the method developed by Devchand and 
Ogievetsky in refs.~\cite{doe,do}.

\subsection{SDSG vielbeine and connections}

Our starting point here is the SDSG constraints (2.35)--(2.38) in a supersymmetric
gauge $\o^+_A=0$. The covariant derivatives in terms of the corresponding
vielbeine and superconnections read in the chiral superspace of the $N$-extended
SDSG as follows (see sect.~2.2):
$$ \nabla^{+}_{A}=e^{+}_{A}=E_{A}^{M}\partial^{+}_{M}~,\eqno(4.1)$$
$$ \nabla^{-}_{A}=e^{-}_{A}+\omega^{-}_{A}
=F_{A}^{M}\partial^{-}_{M}+F_{A}^{--M}\partial^{+}_{M}+\omega^{-}_{A}~,
\eqno(4.2)$$
$$ \nabla^{++}=D^{++}+\omega^{++}
=H^{+M}\partial^{+}_{M}+H^{+3M}\partial^{-}_{M}+\omega^{++}~.\eqno(4.3)$$
Substituting eq.~(4.1) into the left-hand-side of eq.~(2.35) yields
$$\eqalign{
\[ \nabla^{+}_{A},\nabla^{+}_{B} \} ~=  & \[ E_{A}^{M}\partial^{+}_{M},
E_{B}^{N}\partial^{+}_{N} \} \cr
~= & E_{A}^{N}(\partial^{+}_{N}E_{B}^{M})\partial^{+}_{M}
-(-)^{AB}E_{B}^{N}(\partial^{+}_{N}E_{A}^{M})\partial^{+}_{M}~,\cr}
$$
where the sign factor $(-)^{AB}$ means the standard grading. Thus we obtain
$$ E_{A}^{N}(\partial^{+}_{N}E_{B}^{M})
-(-)^{AB}E_{B}^{N}(\partial^{+}_{N}E_{A}^{M})=0~,\eqno(4.4)$$
which is nothing but the holonomy condition for the vielbein $e^+_A$. Hence,
there exists a gauge in which this vielbein is holonomic (see subsect.~4.2).

Substituting eqs.~(4.1) and (4.3) into the left-hand-side of eq.~(2.36) yields
$$\eqalign{
\[ \nabla^{++},\nabla^{+}_{A} \] ~=& \[ D^{++}+\omega^{++},e^{+}_{A} \] \cr
~=& \[ D^{++},e^{+}_{A} \] + \[ \omega^{++},e^{+}_{A} \]~,\cr}\eqno(4.5)$$
where
$$ \[ D^{++},e^{+}_{A} \] =
D^{++}E_{A}^{M}\partial^{+}_{M}
-D^{+}_{A}H^{+M}\partial^{+}_{M}+D^{+}_{A}H^{+3M}\partial^{-}_{M}
\eqno(4.6)$$
and
$$\eqalign{
\[ \omega^{++},e^{+}_{A} \] ~=& \[ \omega^{++}_{CD}M^{DC},e^{+}_{A} \] \cr
~=& \omega^{++}_{CD}(M^{DC}e^{+}_{A})-e^{+}_{A}(\omega^{++}_{CD}M^{DC})\cr
~=& \omega^{++}_{CD}(M^{DC}e^{+}_{A})
-D^{+}_{A}\omega^{++}-(-)^{A(C+D)}\omega^{++}_{CD}(e^{+}_{A}M^{DC})\cr
~=& \omega^{++}_{CD}\[ M^{DC},e^{+}_{A} \}
-D^{+}_ {A}\omega^{++}~.\cr}\eqno(4.7)$$
Eqs.~(2.16) and (4.7) now imply
$$\eqalign{
\[ \omega^{++},e^{+}_{A} \] ~=&
2\omega^{++}_{AC}\eta^{DC}e^{+}_{D}-D^{+}_{A}\omega^{++}\cr
~=& 2\omega^{++}_{AC}\eta^{DC}E_{D}^{M}\partial^{+}_{M}-D^{+}_{A}\omega^{++}
~.\cr}\eqno(4.8)$$
Putting our results together, we find for the left-hand-side of eq.~(2.36)
$$ \eqalign{
\[ \nabla^{++},\nabla^{+}_{A} \] ~= &
D^{++}E_{A}^{M}\partial^{+}_{M}-D^{+}_{A}H^{+M}\partial^{+}_{M}
-D^{+}_{A}H^{+3M}\partial^{-}_{M}\cr
~& + 2\omega^{++}_{AC}\eta^{DC}E_{D}^{M}\partial^{+}_{M}
-D^{+}_{A}\omega^{++}~.\cr}\eqno(4.9)$$
Hence, all the coefficients in front of $\partial^{+}_{M}$ and $\partial^{-}_{M}$,
as well as the connection term in eq.~(4.9) should vanish. The first vanishing
coefficient
$$ D^{++}E_{A}^{M}-D^{+}_{A}H^{+M}+2\omega^{++}_{AC}\eta^{DC}E_{D}^{M}=0
\eqno(4.10)$$
determines the harmonic connection $\o^{++}$ in terms of the other fields. The
second vanishing coefficient
$$ D^{+}_{A}H^{+3M}=0 \eqno(4.11)$$
means that the harmonic vielbein  $H^{+3M}$ is analytic,
$$ H^{+3M}=H^{+3M}(z^{+},u)~.\eqno(4.12)$$
The rest of eq.~(4.9) gives the equation
$$ D^{+}_{A}\omega^{++}=0~,\eqno(4.13)$$
whose solution means that the harmonic connection $\o^{++}$ is an analytic
superfield too,
$$ \omega^{++}=\omega^{++}(z^{+},u)~.\eqno(4.14)$$

Sunstituting eqs.~(4.1) and (4.2) into eq.~(2.37) yields
$$\eqalign{
\[ \nabla^{+}_{A},\nabla^{-}_{B} \} ~=&
\[ e^{+}_{A},e^{-}_{B}+\omega^{-}_{B} \} \cr
~=& \[ e^{+}_{A},e^{-}_{B} \} + \[ e^{+}_{A},\omega^{-}_{B} \}~,\cr}
\eqno(4.15)$$
where
$$\[ e^{+}_{A},e^{-}_{B} \}
= D^{+}_{A}F_{B}^{M}\partial^{-}_{M}+D^{+}_{A}F_{B}^{--M}\partial^{+}_{M}
-(-)^{AB}D^{-}_{B}E_{A}^{M}\partial^{+}_{M}~,\eqno(4.16)$$
and
$$\eqalign{
\[ e^{+}_{A},\omega^{-}_{B} \} ~=&
e^{+}_{A}(\omega^{-}_{BCD}M^{DC})-(-)^{AB}\omega^{-}_{BCD}(M^{DC}e^{+}_{A})\cr
~=& D^{+}_{A}\omega^{-}_{B}+(-)^{A(B+C+D)}\omega^{-}_{BCD}(e^{+}_{A}M^{DC})
-(-)^{AB}\omega^{-}_{BCD}(M^{DC}e^{+}_{A})\cr
~=& D^{+}_{A}\omega^{-}_{B}-(-)^{AB}\omega^{-}_{BCD}\[ M^{DC},e^{+}_{A} \}\cr
~=& D^{+}_{A}\omega^{-}_{B}-(-)^{AB}2\omega^{-}_{BAC}\eta^{DC}e^{+}_{D}\cr
~=& D^{+}_{A}\omega^{-}_{B}-(-)^{AB}2\omega^{-}_{BAC}\eta^{DC}E_{D}^{M}
\partial^{+}_{M}~.\cr}\eqno(4.17)$$
Putting this together, one finds
$$\eqalign{
\[ \nabla^{+}_{A},\nabla^{-}_{B} \} ~=&
D^{+}_{A}F_{B}^{M}\partial^{-}_{M}+D^{+}_{A}F_{B}^{--M}\partial^{+}_{M}
-(-)^{AB}D^{-}_{B}E_{A}^{M}\partial^{+}_{M}\cr
~+& D^{+}_{A}\omega^{-}_{B}-(-)^{AB}2\omega^{-}_{BAC}\eta^{DC}E_{D}^{M}
\partial^{+}_{M}~.\cr}\eqno(4.18)$$
Comparing the terms having $\pa^+_M$ with the right-hand-side of eq.~(2.37) yields
$$ D^{+}_{A}F_{B}^{--M}-(-)^{AB}D^{-}_{B}E_{A}^{M}-(-)^{AB}2\omega^{-}_{BAC}
\eta^{DC}E_{D}^{M}=0~,\eqno(4.19)$$
which is an algebraic equation for the connection $\omega^{-}_{A}$ in terms of
the vielbeine. Similarly, when comparing the terms having $\partial^{-}_{M}$, one
finds
$$ D^{+}_{A}F_{B}^{M}=0~.\eqno(4.20)$$
Hence, the vielbein $F_{A}^{M}$ is also analytic,
$$ F_{A}^{M}=F_{A}^{M}(z^{+},u)~.\eqno(4.21)$$

Comparing the rest of eq.~(4.18) with the right-hand-side of eq.~(2.37) yields
$$ D^{+}_{A}\omega^{-}_{B}=R_{AB}~,\eqno(4.22)$$
which simply defines the super-Riemann tensor $R_{AB}$.

Finally, after substituting eqs.~(4.2) and (4.3) into eq.~(2.38), one gets
$$\eqalign{
\[ \nabla^{++},\nabla^{-}_{A} \] ~=& \[ D^{++}+\omega^{++},
e^{-}_{A}+\omega^{-}_{A} \] \cr
~=& \[ D^{++},e^{-}_{A} \] + \[ D^{++},\omega^{-}_{A} \] +\[ \omega^{++},
e^{-}_{A} \] + \[ \omega^{++},\omega^{-}_{A} \]~,\cr}\eqno(4.23)$$
where
$$ \[ D^{++},e^{-}_{A} \]= D^{++}F_{A}^{M}\partial^{-}_{M}
+D^{++}F_{A}^{--M}\partial^{+}_{M}
-D^{-}_{A}H^{+M}\partial^{+}_{M}
-D^{-}_{A}H^{+3M}\partial^{-}_{M}~,\eqno(4.24)$$
$$  \[ D^{++},\omega^{-}_{A} \] = D^{++}\omega^{-}_{A}~,\eqno(4.25)$$
and
$$\eqalign{
\[ \omega^{++},e^{-}_{A} \] ~=&
\omega^{++}_{CD}(M^{DC}e^{-}_{A})-e^{-}_{A}(\omega^{++}_{CD}M^{DC})\cr
~=& \omega^{++}_{CD}(M^{DC}e^{-}_{A})-D^{-}_{A}\omega^{++}
-(-)^{A(C+D)}\omega^{++}_{CD}(e^{-}_{A}M^{DC})\cr
~=& \omega^{++}_{CD}\[ M^{DC},e^{-}_{A} \}
-D^{-}_{A}\omega^{++}\cr
~=& 2\omega^{++}_{AC}\eta^{DC}e^{-}_{D}-D^{-}_{A}\omega^{++}\cr
~=& 2\omega^{++}_{AC}\eta^{DC}F_{D}^{M}\partial^{-}_{M}+2\omega^{++}_{AC}
\eta^{DC}F_{D}^{--M}\partial^{+}_{M}-D^{-}_{A}\omega^{++}~.\cr}\eqno(4.26)$$

Putting all the terms together on the right-hand-side of eq.~(2.38) yields
$$
D^{++}F_{A}^{M}\partial^{-}_{M}+D^{++}F_{A}^{--M}\partial^{+}_{M}-D^{-}_{A}
H^{+M}\partial^{+}_{M}-D^{-}_{A}H^{+3M}\partial^{-}_{M}-D^{-}_{A}\omega^{++}$$
$$+2\omega^{++}_{AC}\eta^{DC}F_{D}^{M}\partial^{-}_{M}+2\omega^{++}_{AC}\eta^{DC}
F_{D}^{--M}\partial^{+}_{M}+D^{++}\omega^{-}_{A}+\[ \omega^{++},\omega^{-}_{A}
\] = E_{A}^{M}\partial^{+}_{M}~.$$
Hence, when comparing the terms with $\partial^{+}_{M}$, one finds that
$$
D^{++}F_{A}^{--M}+2\omega^{++}_{AC}\eta^{DC}F_{D}^{--M}-D^{-}_{A}H^{+M}
=E_{A}^{M}~.\eqno(4.27)$$
This equation gives the vielbein $E_{A}^{M}$ in terms of the other superfields.
As we already mentioned at the beginning of this subsection, this vielbein can
be put into a holonomic form via a proper coordinate transformation. Eq.~(4.27)
simultaneously gives an algebraic relation between $E_{A}^{M}$ and $\o^{++}$.
Since the connection $\o^{++}$ was already expressed in terms of the vielbeine 
in eq.~(4.10), eqs.~(4.10) and (4.27) together imply a first-order 
{\it differential} equation of motion for the vielbeine.

Comparing the coefficients in front of $\pa^-_M$ on the right-hand-side of
eq.~(2.38) yields the relation
$$ D^{++}F_{A}^{M}+2\omega^{++}_{AC}\eta^{DC}F_{D}^{M}-D^{-}_{A}H^{+3M}=0~.
\eqno(4.28)$$
Similarly, since the vielbein $F_{A}^{M}$ can also be put into a holonomic form
(see the next subsect.~4.2), whereas the connection $\omega^{++}$ is now a 
certain function of the vielbeine and their first-order derivatives, eq.~(4.28) 
is actually a first-order {\it differential} equation of motion for the 
harmonic vielbeine $H^{+M}$ and $H^{+3M}$. 
Its solution will be given in the next section.

Finally, the rest of the last SDSG constraint (2.38) gives the equation
$$ D^{++}\omega^{-}_{A}-D^{-}_{A}\omega^{++} +\[ \omega^{++},\omega^{-}_{A}\]=0~,
\eqno(4.29)$$
which is analogous to the first SDSYM equation of motion in eq.~(3.19). 

In this subsection we reduced the number of the relevant SDSG constraints and 
that of the relevant superfields in the on-shell superspace description
of SDSG. However, despite of the {\it Ansatz\/} of harmonic superspace in the 
analytic coordinates, the obtained system of SDSG equations is still rather 
complicated, and it has some redundant gauge symmetries acting on superfields. 
An important next step will be the introducion of a Frobenius gauge~\cite{do}, 
which will result is a great simplification of the superfield SDSG.

\subsection{Frobenius gauge}

The vielbeine $(E^M_A,F^M_A,F^{--M}_A,H^{+3M},H^{+M})$ are subject to the
superspace diffeomorphisms with the infinitesimal analytic superfield parameters
$(\l^{+M},\l^{-M})$, and the local $OSp(N|2)$ super-Lorentz rotations with the
inifinitesimal analytic superfield parameters $\l^{A}_{B}$, whose form is given
by eq.~(2.56). These gauge symmetries can be used to impose a supersymmetric gauge
which would simplify the SDSG constraints.

Since the covariant derivatives $\de^+_A$ (anti)commute,
$$ \[\nabla^{+}_{A},\nabla^{+}_{B}\}=0~,\eqno(4.30)$$
it follows from the Frobenius theorem that there exists a coordinate system in
which $\de^+_A=\pa^+_A$. In addition, because of the analyticity of $F_{A}^{M}$
(see eq.~(4.20)), there exists an $OSp(N|2)$ rotation which brings the supermatrix
$F_{A}^{M}$ to a unit supermatrix. This leads to the Frobenius gauge~\cite{do}
$$  E_{A}^{M}=\delta_{A}^{M}~,\quad F_{A}^{M}=\delta_{A}^{M}~.\eqno(4.31)$$
It is obvious that in this gauge there is no difference between the world and
tangent superspace indices. The SDSG dynamics is described in terms of the
constrained superfields $H^{+3M},H^{+M},F_A^{--M},~\o^{++}$ and $\o_A^-$~.

Let us now investigate which residual symmetries survive in the gauge (4.31).
It follows from $\d F\du{A}{M}=0$ that
$$\eqalign{
0~=&~\delta_{A}^{N}\partial^{-}_{N}\lambda^{M+}+\lambda_{A}^{B}\delta_{B}^{M} \cr
~=&~\partial^{-}_{A}\lambda^{B+}\delta_{B}^{M}+\lambda_{A}^{B}\delta_{B}^{M}\cr
~=&~\partial^{-}_{A}\lambda^{B+}+\lambda_{A}^{B}~.\cr}\eqno(4.32)$$
After taking the supertrace of eq.~(4.32) and using the supertracelessness of the
$OSp(N|2)$ parameters $\l^B_A$, one easilty finds that the analytic
diffeomorphism parameters $ \lambda^{\pm B}(z^{+},u)$ satisfy
$$ \partial^{-}_{A}\lambda^{A+}=0~.\eqno(4.33)$$
A solution to this equation is given by
$$ \lambda^{+B}(z^{+},u)=\partial^{-}_{C}\lambda^{++}(z^{+},u)\eta^{BC}~.
\eqno(4.34)$$
Because of eqs.~(4.32) and (4.34), the super-Lorentz parameters $\lambda_{A}^{B}$
are now dependent upon the just introduced parameter $\lambda^{++}(z^+,u)$ as
follows:
$$\eqalign{
\lambda_{A}^{B}~=~&-\partial^{-}_{A}\lambda^{+B}(z+,u)\cr
~=~&-\partial^{-}_{A}\partial^{-}_{C}\lambda^{++}\eta^{BC}~.\cr}\eqno(4.35)$$

Similarly, it follows from $\d E^M_A=0$ that
$$ \eqalign{
0~=~&\delta_{A}^{N}\partial^{+}_{N}\lambda^{M-}+\lambda_{A}^{B}\delta_{B}^{M}\cr
~=~&\partial^{+}_{A}\lambda^{B-}\delta_{B}^{M}+\lambda_{A}^{B}\delta_{B}^{M}\cr
~=~&\partial^{+}_{A}\lambda^{B-}-\partial^{-}_{A}\partial^{-}_{C}\lambda^{++}
\eta^{BC}~.\cr}\eqno(4.36)$$
A solution to this equation reads
$$ \lambda^{B-}=z^{D-}\partial^{-}_{D}\partial^{-}_{C}\lambda^{++}\eta^{BC}
+\tilde\lambda^{B-}(z^{+},u)~,\eqno(4.37)$$
where $\tilde\lambda^{-B}(z^{+},u)$ is an arbitrary analytic function. In order
to check eq.~(4.37), it is enough to notice that
$$\eqalign{
\partial^{+}_{A}\lambda^{B-}~=~&\partial_{A}^{+}(z^{D-}\partial^{-}_{D}
\partial^{-}_{C}\lambda^{++})\eta^{BC}+\partial^{+}_{A}\tilde\lambda^{B-}\cr
~=~&\delta_{A}^{D}\partial^{-}_{D}\partial^{-}_{C}\lambda^{++}\eta^{BC}\cr
~=~&\partial^{-}_{A}\partial^{-}_{C}\lambda^{++}\eta^{BC}~.\cr}\eqno(4.38)$$

Therefore, the residual symmetries of the Frobenius gauge are described in terms
of the unconstrained superfield parameters $\lambda^{++}(z^{+},u)$ and 
$\tilde\lambda^{-B}(z^{+},u)$ introduced above. We are now in a position to 
define superfield prepotentials of SDSG in the Frobenius gauge.

\section{SDSG prepotentials in the Frobenius gauge}

Let us now investigate the SDSG equations of subsect.~(4.1) in the
Frobenius gauge (4.31).

Eq.~(4.10) reduces in the Frobenius gauge to
$$ \partial^{+}_{A}H^{+M}=2\omega^{++}_{AC}\eta^{DC}\delta_{D}^{M}~,
\eqno(5.1)$$
which determines the connection $\o^{++}$ in terms of the vielbein $H^{M+}$.
Since $\o^{++}$ is analytic (see eq.~(4.13)), i.e. $ \pa^+_A\o^{++}=0$,
the superfield $\partial^{+}_{A}H^{+M}$ should therefore be analytic also.

Taking into account eq.~(4.11) in the Frobenius gauge,
$\partial^{+}_{A}H^{+3M}=0$, eq.~(4.28) goes over to
$$\partial^{-}_{A}H^{+3M}=2\omega^{++}_{AC}\eta^{DC}\delta_{D}^{M}~,\eqno(5.2)$$
where we have used the relations
$$ D^{-}_{A}H^{+3M}=\partial^{-}_{A}H^{+3M}+F_{A}^{--N}\partial^{+}_{N}H^{+3M}
=\partial^{-}_{A}H^{+3M}~.\eqno(5.3)$$
Comparing eqs.~(5.1) and (5.2) obviously yields
$$ \partial^{+}_{A}H^{+M}=\partial^{-}_{A}H^{+3M}~,\eqno(5.4)$$
whose solution is
$$ H^{+M}=z^{-N}\partial^{-}_{N}H^{+3M}~.\eqno(5.5)$$

The constraint (5.2) can be solved in terms of an arbitrary superfield
$H^{+4}(z^+,u)$ as follows:
$$\eqalign{
2\omega^{++}_{AC}\eta^{DC}\delta_{D}^{M} ~=~& \partial^{-}_{A}H^{+3M}\cr
~=~& \partial^{-}_{A}H^{+3}_{C}\eta^{DC}\delta_{D}^{M}\cr
~=~& \partial^{-}_{A}\partial^{-}_{C}H^{+4}\eta^{DC}\delta_{D}^{M}~,\cr}
\eqno(5.6)$$
or, equivalently, 
$$ \omega^{++}_{AC}=\frac{1}{2}\partial^{-}_{A}\partial^{-}_{C}H^{+4} ~,
\eqno(5.7)$$
and
$$ H^{+3M}=\partial^{-}_{C}H^{+4}\eta^{DC}\delta_{D}^{M}~.\eqno(5.8)$$
The vielbeine and connections, which are associated with the harmonic derivative
$\de^{++}$,  can therefore be entirely expressed in terms of the single analytic
prepotential $H^{+4}(z^+,u)$  of charge $(+4)$~\cite{do}.

Eq.~(4.19) in the Frobenius gauge reduces to
$$
2\omega^{-}_{BAC}\eta^{DC}\delta_{D}^{M}=(-)^{AB}\partial^{+}_{A}F_{B}^{--M}~,
\eqno(5.9)$$
which is similar to eq.~(5.1). Our {\it Ansatz} for its solution, in terms of 
an independent superfield $V^{-4}$ of charge $(-4)$, reads
$$
F_{B}^{--M}=\partial^{+}_{B}\partial^{+}_{C}V^{-4}\eta^{DC}\delta_{D}^{M}~.
\eqno(5.10)$$
Substituting eq.~(5.10) into eq.~(5.9) yields
$$\eqalign{
2\omega^{-}_{BAC}\eta^{DC}\delta_{D}^{M} ~=~&
(-)^{AB}\partial^{+}_{A}\partial^{+}_{B}\partial^{+}_{C}V^{-4}\eta^{DC}
\delta_{D}^{M}\cr
~=~& \partial^{+}_{B}\partial^{+}_{A}\partial^{+}_{C}V^{-4}\eta^{DC}
\delta_{D}^{M}~,\cr}\eqno(5.11)$$
and, hence,
$$ \omega^{-}_{BAC}=\frac{1}{2}\partial^{+}_{B}\partial^{+}_{A}\partial^{+}_{C}
V^{-4}~.\eqno(5.12)$$
Unlike the analytic prepotential $H^{+4}$, the superfield  $V^{-4}$ is, however, 
not analytic.

The only remaining equations (4.27) and (4.29) in the Frobenius gauge take the
form
$$ D^{++}F_{A}^{--M}+2\omega^{++}_{AC}\eta^{DC}F_{D}^{--M}-D^{-}_{A}H^{+M}
=\delta_{A}^{M}~,\eqno(5.13)$$
and (unchanged)
$$ D^{++}\omega^{-}_{A}-D^{-}_{A}\omega^{++}
+\[ \omega^{++},\omega^{-}_{A} \]=0~,\eqno(5.14)$$
respectively. When considering eq.~(5.13) as an equation on $V^{-4}$, one might
solve it (at least, in principle) in terms of the analytic prepotential $H^{+4}$.
The last equation (5.14) might then be considered as a single equation of motion.
However, these two equations (5.13) and (5.14) are actually not independent. 
Indeed, given eq.~(5.13), eq.~(5.14) is automatically satisfied, which becomes 
obvious after differentiating eq.~(5.13) with respect to $\pa^+_B$.
This means that, via a solution of eq.~(5.13) in terms of an arbitrary
prepotential $H^{+4}$, the SDSG theory is automatically on-shell. Therefore,
one arrives at the formal general solution to the SDSG equations via the twistor
transform, in terms of an arbitrary analytic superfield $H^{+4}(z^+,u)$,
as in ref.~\cite{do}. Of course, this solution is still formal, since its
relation to the initial central basis is given via the set of linear differential
equations on the `bridge' functions, {\it viz.}
$$ H^{+3M}=\partial^{++}b^{M+}~,\quad
H^{+M}= z^{+M}+\partial^{++}b^{-M}~,\quad
\omega^{++}=\vf^{-1}D^{++}\vf~,\eqno(5.15)$$
whose manifest solutions are required in order to express the component vielbeine
in terms of the component prepotentials.

In order to write down an off-shell action, we are not allowed to use the
analytic prepotential $H^{+4}$ since it exists only on-shell. Therefore, we 
should relax our constraints first, which will be described in the next 
sect.~6.

\section{An action of ~$N=8$~ SDSG}

It follows from eq.~(2.38) that the covariant derivative $\de^-_A$ in the 
central coordinates is linear in harmonics $u^-$. This implies 
(see subsects.~2.2 and 2.3) that
$$ \[ \nabla^{-}_{A},\nabla^{-}_{B} \} =0~.\eqno(6.1)$$
The left-hand-side of this equation,
$$ \eqalign{
\[ \nabla^{-}_{A},\nabla^{-}_{B} \} ~=~& 
\[ e^{-}_{A}+\omega^{-}_{A},e^{-}_{B}+\omega^{-}_{B} \} \cr
~=~& \[ e^{-}_{A},e^{-}_{B} \} + \[ e^{-}_{A},\omega^{-}_{B} \} 
+ \[ \omega^{-}_{A},e^{-}_{B} \} + \[ \omega^{-}_{A},\omega^{-}_{B} \}~,
\cr}\eqno(6.2)$$
can be written down in a more explicit form by using the equations
$$ \[ e^{-}_{A},e^{-}_{B} \} 
= D^{-}_{A}F_{B}^{--M}\partial^{+}_{M}-(-)^{AB}D^{-}_{B}F_{A}^{--M}
\partial^{+}_{M}~,\eqno(6.3)$$
$$ \eqalign{
\[ e^{-}_{A},\omega^{-}_{B}\}~=~& e^{-}_{A}(\omega^{-}_{BCD}M^{DC})
-(-)^{AB}\omega^{-}_{BCD}(M^{DC}e^{-}_{A})\cr
~=~& D^{-}_{A}\omega^{-}_{B}+(-)^{A(B+C+D)}\omega^{-}_{BCD}(e^{-}_{A}M^{DC})
-(-)^{AB}\omega^{-}_{BCD}(M^{DC}e^{-}_{A})\cr
~=~& D^{-}_{A}\omega^{-}_{B}-(-)^{AB}\omega^{-}_{BCD} \[ M^{DC}e^{-}_{A}
-(-)^{A(C+D)}e^{-}_{A}M^{DC} \} \cr
~=~& D^{-}_{A}\omega^{-}_{B}-(-)^{AB}\omega^{-}_{BCD}\[ M^{DC},e^{-}_{A} \}\cr
~=~& D^{-}_{A}\omega^{-}_{B}-(-)^{AB}2\omega^{-}_{BAC}\eta^{DC}e^{-}_{D}\cr
~=~& D^{-}_{A}\omega^{-}_{B}-(-)^{AB}2\omega^{-}_{BAC}\eta^{DC}\delta_{D}^{M}
\partial^{-}_{M}
-(-)^{AB}2\omega^{-}_{BAC}\eta^{DC}F_{D}^{--M}\partial^{+}_{M}~,\cr}
\eqno(6.4)$$
and
$$\eqalign{
\[ \omega^{-}_{A},e^{-}_{B} \}~=~& -(-)^{AB}D^{-}_{B}\omega^{-}_{A}
+2\omega^{-}_{BAC}\eta^{DC}e^{-}_{D}\cr
~=~& -(-)^{AB}D^{-}_{B}\omega^{-}_{A}+2\omega^{-}_{ABC}\eta^{DC}\delta_{D}^{M}
\partial^{-}_{M}+2\omega^{-}_{ABC}\eta^{DC}F_{D}^{--M}\partial^{+}_{M}~.\cr}
\eqno(6.5)$$
Putting all the equations together into eq.~(6.1) and comparing the 
coefficients in front of the derivative $\pa^+_M$ yields
$$ D^{-}_{A}F_{B}^{--M}-(-)^{AB}D_{B}^{-}F_{A}^{--M}-(-)^{AB}2\omega^{-}_{BAC}
\eta^{DC}F_{D}^{--M}$$
$$ +\,2\omega^{-}_{ABC}\eta^{DC}F_{D}^{--M}=0~.\eqno(6.6)$$
Similarly, the terms with $\pa^-_M$ imply the relation
$$ \omega^{-}_{ABC}=(-)^{AB}\omega^{-}_{BAC}~.\eqno(6.7)$$
Eq.~(6.7) allows us to simplify eq.~(6.6) to the form
$$ D^{-}_{A}F_{B}^{--M}-(-)^{AB}D_{B}^{-}F_{A}^{--M}=0~.\eqno(6.8)$$

Comparing the terms without derivatives in eq.~(6.1) adds the equation 
$$ D^{-}_{A}\omega^{-}_{B}-(-)^{AB}D^{-}_{B}\omega^{-}_{A}+\[
 \omega^{-}_{A},\omega^{-}_{B} \}=0~.\eqno(6.9)$$
It is easy to verify that differentiating eq.~(6.8) with respect to $\pa^+_A$
yields eq.~(6.9). This means that eq.~(6.9) is not independent and can be 
ignored, while eq.~(6.8) represents the true equation of motion. In terms of the 
prepotential $V^{-4}$ defined by eq.~(5.10),~\footnote{A similar prepotential was
introduced in ref.~\cite{doe}, in the case of the purely bosonic SDG.} eq.~(6.8) 
takes the form
$$(\partial_{A}^{-}\partial_{B}^{+}-\partial_{A}^{+}\partial_{B}^{-})
\partial_{C}^{+} V^{-4}+\partial_{A}^{+}\partial_{E}^{+}V^{-4}\eta^{FE}
\partial_{F}^{+}\partial_{B}^{+}\partial_{C}^{+}V^{-4}$$
$$-(-)^{AB}\partial_{B}^{+}\partial_{E}^{+}V^{-4}\eta^{FE}\partial_{F}^{+}
\partial^{+}_{A}\partial_{C}^{+}V^{-4}=0~.\eqno(6.10)$$

When using the graded Leibniz-rule as well as the graded antisymmetry of the
metric tensor $\eta^{AB}$, one easily finds from eq.~(6.10) that
$$ \partial^{+}_{C}\left[ 
(\partial_{A}^{-}\partial^{+}_{B}-\partial^{+}_{A}\partial^{-}_{B})V^{-4}
+(\partial^{+}_{A}\partial^{+}_{D}V^{-4})\eta^{ED}
(\partial^{+}_{E}\partial^{+}_{B}V^{-4})\right]=0~.\eqno(6.11)$$
This simply means that the function in the rectangular brackets is analytic
and, therefore, it can be gauged away to zero by using the invariance of the
defining eq.~(5.10) with respect to the pre-gauge transformations of the 
prepotential,
$$ \delta V^{-4}=z^{-A}\L_{A}^{-3}(z^{+},u)+\L^{-4}(z^{+},u)~,\eqno(6.12)$$
with the infinitesimal analytic superfield parameters $\L^{-3}_A$ and 
$\L^{-4}$.
Therefore, the equation of motion in terms of the prepotential $V^{-4}$ takes
the form 
$$(\partial^{-}_{A}\partial^{+}_{B}-\partial^{+}_{A}\partial^{-}_{B})V^{-4}
+(\partial^{+}_{A}\partial^{+}_{D}V^{-4})\eta^{ED}(\partial^{+}_{E}
\partial^{+}_{B}V^{-4})=0~.\eqno(6.13)$$
In this form, it is very similar to the Siegel equation of motion, which 
follows from the light-cone action (2.25). Our equation of motion is, however,
Lorentz-covariant unlike that of Siegel.

Eq.~(6.13) amounts to the following equations
$$\eqalign{
(\partial^{-}_{\alpha}\partial^{+}_{\beta}-\partial^{+}_{\alpha}
\partial^{-}_{\beta})V^{-4}+(\partial^{+}_{\alpha}\partial^{+}_{D}V^{-4})
\eta^{ED}(\partial^{+}_{E}\partial^{+}_{\beta}V^{-4})~=&0~,\cr
(\partial^{-}_{a}\partial^{+}_{b}-\partial^{+}_{a}\partial^{-}_{b})V^{-4}
+(\partial^{+}_{a}\partial^{+}_{D}V^{-4})\eta^{ED}(\partial^{+}_{E}
\partial^{+}_{b}V^{-4})~=~&0~,\cr
(\partial^{-}_{\alpha}\partial^{+}_{b}-\partial^{+}_{\alpha}\partial^{-}_{b})
V^{-4}+(\partial^{+}_{\a}\partial^{+}_{D}V^{-4})\eta^{ED}(\partial^{+}_{E}
\partial^{+}_{b}V^{-4})~=~&0~,\cr}
\eqno(6.14)$$
where we have simply used the fact that $A=(a,\a)$ and $B=(b,\b)$. The 
Grassmann derivatives $\pa^-_a$ enter the second and third lines of eq.~(6.14)
 only linearly. Hence, as in ref.~\cite{sie}, we can use these two equations
in order  to solve all the `non-analytic' (i.e. $\q^{+}$-dependent) terms in the
prepotential $V^{-4}$ as the functions of the remaining `anti-analytic' 
components. This leads to the effectively anti-analytic prepotential 
$\left.V^{-4}\right|_{\q^{+}=0}\,$. This statement 
is, in fact, basis-independent since the expansion rules  with respect to the 
anicommuting superspace coordinates in any basis are all isomorphic.

Expanding the $N=8$ prepotential in terms of the anticommuting anti-analytic 
coordinates yields
$$\eqalign{
V^{-4}(x,\theta^{-},u)~=~&e^{-4}+\psi^{-3}_{a}\theta^{-a}+A^{-2}_{[ab]}
\theta^{-2ab} +\chi^{-}_{[abc]}\theta^{-3abc}+\phi_{[abcd]}\theta^{-4abcd}\cr
~+~&\chi^{+[abc]}\theta^{-5}_{abc}+A^{+2[ab]}\theta^{-6}_{ab}+\psi^{+3a}
\theta^{-7}_{a}+e^{+4}\theta^{-8}~,\cr}\eqno(6.15)$$
where we have introduced the notation
$$ \eqalign{
\theta^{-a}~=~&\theta^{-a}~,\cr
\theta^{-2ab}~=~&\frac{1}{2!}\theta^{-a}\theta^{-b}~,\cr
\theta^{-3abc}~=~&\frac{1}{3!}\theta^{-a}\theta^{-b}\theta^{-c}~,\cr
\theta^{-4abcd}~=~&\frac{1}{4!}\theta^{-a}\theta^{-b}\theta^{-c}\theta^{-d}
~,\cr
\theta^{-5}_{abc}~=~&\frac{1}{5!}\epsilon_{abcdefgh}
\theta^{-d}\theta^{-e}\theta^{-f}\theta^{-g}\theta^{-h}~,\cr
\theta^{-6}_{ab}~=~&\frac{1}{6!}\epsilon_{abcdefgh}\theta^{-c}
\theta^{-d}\theta^{-e}\theta^{-f}\theta^{-g}\theta^{-h}~,\cr
\theta^{-7}_{a}~=~&\frac{1}{7!}\epsilon_{abcdefgh}\theta^{-b}
\theta^{-c}\theta^{-d}\theta^{-e}\theta^{-f}\theta^{-g}\theta^{-h}~,\cr
\theta^{-8}~=~&\frac{1}{8!}\epsilon_{abcdefgh}\theta^{-a}\theta^{-b}
\theta^{-c}\theta^{-d}\theta^{-e}\theta^{-f}\theta^{-g}\theta^{-h}~.\cr}
\eqno(6.16)$$
The charges of $2^N$ component fields appearing in the expansion (6.15) can
be identified with the component `helicities' to be multiplied by a factor $2$. 
The prepotential $V^{-4}$ is obviously self-adjoint in the case of $N=8$ only.

Multiplying the last remaining equation of motion in the first line of
eq.~(6.14) with $C^{\a\b}$ yields
$$ \bo V^{-4}+\frac{1}{2}(\partial^{+\alpha}\partial^{+}_{D}V^{-4})
\eta^{ED}(\partial^{+}_{E}\partial^{+}_{\alpha}V^{-4})=0~,\eqno(6.17)$$
where we have introduced the d'Alembertian
$\bo=(\partial^{-}_{\alpha}\partial^{+}_{\beta}
-\partial^{+}_{\alpha}\partial^{-}_{\beta})C^{\alpha\beta}$. 

The action, whose equations of motion are given by eq.~(6.17) in the case of
the maximally extended $N=8$ SDSG, reads 
$$ S=\int d^{4}xd^{+8}\theta du\, \left\{
\frac{1}{2}V^{-4}\bo V^{-4} + \frac{1}{6}V^{-4}(\partial^{+\alpha}
\partial^{+}_{D}V^{-4})\eta^{ED}(\partial^{+}_{E}\partial^{+}_{\alpha}
V^{-4})\right\}~.\eqno(6.18)$$

The action (6.18) is our main result in this paper. Though a similar action
can be written down for any $N$, it is Lorentz-invariant only in the case of 
$N=8$ SDSG, whose measure is dimensionless and of charge $(+8)$. The action
is also invariant under the residual gauge transformations in the Frobenius
gauge. In the analytic coordinates, they are given by the analytic 
difeomorphisms whose infinitesimal superfield parameters are
$$
\lambda^{+A}(z^{+},u)=\partial^{-}_{B}\lambda^{++}(z^{+},u)\eta^{AB}~,
\eqno(6.19)$$
and
$$ \lambda^{-A}(z^{\pm},u)=z^{-C}\partial^{-}_{C}\partial^{-}_{B}
\lambda^{++}(z^{+},u)\eta^{AB} +\tilde\lambda^{-A}(z^{+},u)~,\eqno(6.20)$$
and the analytic $OSp(8|2)$ rotations whose infinitesimal superfield 
parameters are
$$ \lambda_{A}^{B}=-\partial^{-}_{A}\partial^{-}_{C}
\lambda^{++}(z^{+},u)\eta^{BC}~,\eqno(6.21)$$
in terms of the independent analytic parameters $\lambda^{++}(z^+,u)$ and 
$\tilde\lambda^{-A}(z^+,u)$ to be evaluated at $\theta^{+}=0$.

\section{Conclusion}

To conclude, let us briefly summarize what we did in the preceeding sections. 
The on-shell $N$-extended SDSG superspace constraints were reformulated in 
terms of the analytic coordinates in harmonic superspace. Then the Frobenius 
gauge was introduced, in which the difference between the world- and tangent- 
superspace indices disappeared, as in ref.~\cite{do}. The harmonic superspace 
constraints were partially solved in terms of two prepotentials $H^{+4}(z^+,u)$
and $V^{-4}(z^+,z^-,u)$. The analytic Devchand-Ogievetsky prepotential $H^{+4}$ 
exists only on-shell, where it formally solves the SDSG equations via the twistor 
transform. In order to describe the theory off-shell, we relaxed the SDSG 
constraints and solved the vanishing graded commutator of the $\de^-_A$ 
covariant derivatives in the Frobenius gauge. Along these lines, no $H^{+4}$
prepotential appears, while the whole theory can be described in terms of 
$V^{-4}$ only ({\it cf.} ref.~\cite{doe}). The equations of motion for the 
prepotential $V^{-4}$ were 
derived and divided into two groups. The first group of equations was used to 
solve all the $\q^{+}$ dependence of the prepotential $V^{-4}$ in terms of the 
remaining anti-analytic components. The only remaiming superfield equation of
motion was then obtained from a Chern-Simons-like harmonic superspace action. 
This $N=8$ SDSG action is very similar to the light-cone $N=8$
SDSG action found by Siegel. It also naturally generalizes the $N=4$ SDSYM
covariant action found by Sokatchev to the case of the $N=8$ SDSG.

Our action (6.18) is manifestly Lorentz-invariant, and it is also manifestly 
supersymmetric with respect to a half of the original supersymmetries of the 
on-shell $N=8$ SDSG by construction. Thus it may be useful e.g., for a study 
of quantum 
properties of the $N=8$ SDSG. We would like to investigate further the meaning 
of the residual gauge symmetries of this action, as well as its hidden 
symmetries. As is well-known, the equations of motion of the non-self-dual 
(ungauged) $N=8$ supergravity have the hidden non-compact global symmetry 
$E_{7(+7)}$~\cite{cj}, which is broken in the $SO(8)$-gauged version of the $N=8$
theory~\cite{wnic}. Nevertheless, even in the gauged $N=8$ supergravity its
$70=133-63$ massless physical scalars can still be considered as the (gauged)
non-linear sigma-model fields taking their values in the target space
$E_7/SU(8)$~\cite{wnic}. Moreover, a discrete subgroup $E_{7(+7)}({\bf Z})$ 
survives as the full $U$-duality group in the four-dimensional (toroidally
compactified) type-II superstring theory which generalizes the non-self-dual
four-dimensional $N=8$ supergravity~\cite{ht}. In particular, the 
$SO(6,6;{\bf Z})$ subgroup of $E_{7(+7)}({\bf Z})$ cab be identified with the
type-II superstring T-duality group. Though the $N=8$ self-dual supergravity 
is much more simpler than the non-self-dual one, the equations of motion of 
the ungauged $N=8$ SDSG seem to have no true duality symmetries beyond the 
manifest $SL(8;{\bf R})$ global rotations. On the other hand, one should 
expect a very rich spectrum of hidden affine-like symmetries in the $N=8$ SDSG
equations of motion, which generalize the known infinite-dimensional 
symmetries of the usual $(N=0)$ SDG~\cite{pof}. The gauged $N=8$ SDSG may be 
relevant for the toroidally compactified four-dimensional heterotic strings 
with a restricted $U$-duality group to be a subgroup of the full $U$-duality 
group $O(6,22;{\bf Z})\otimes SL(2;{\bf Z})$ expected in four 
dimensions~\cite{ht}. The compactified versions of the gauged $N=8$ SDSG 
down to two and three spacetime dimensions may also be relevant for the 
heterotic spinning strings with the $(2,1)$ gauged world-sheet supersymmetry 
and the (compactified) M-theory~\cite{emil}. 

Our analysis does not exclude, in principle, an existence of covariant actions 
for the non-selfconjugate $N<4$ SDSYM and $N<8$ SDSG.~\footnote{See e.g., 
ref.~\cite{bh} for some recent proposals.} However, in order to compensate the
mismatch of `helicities' or $GL(1)'$-charges, any covariant action should
inevitably have an {\it infinite} number of Lagrange multipliers which have to 
compensate each other altogether. These compensating fields are apparently 
different from the auxiliary fields present in the harmonic superfield expansion,
and they may need extra conditions (i.e. beyond the equations of motion which
arise from the action alone) for their actual decoupling on-shell.

\section*{Acknowledgements}

One of the authors (S.V.K.) would like to thank Olaf Lechtenfeld,
Hermann Nicolai and Hitoshi Nishino for useful discussions, and Chand Devchand
for correspondence.

\newpage

\section*{Appendix: $N=8$~ gauged SDSG in components}

The covariant component action of the $N=8$ SDSG was given by Siegel~\cite{sie}. 
It this Appendix we merely reproduce his action for the sake of completeness.
A direct proof of the on-shell equivalence between the Siegel action (A.1) and 
our action (6.18) would require eliminating infinitely many auxiliary fields 
present in the action (6.18) and fixing a gauge in the action (A.1). Though we 
didn't do this off-shell, the on-shell equivalence is nevertheless clear from 
our construction since we started from the same equations of motion in 
superspace. 

The component action was found in ref.~\cite{sie} by extracting the component
equations of motion out of the on-shell $N=8$ SDSG superspace constraints 
(subsect.~2.2) in a covariant (Wess-Zumino) gauge. The independent components
of the $N=8$ SDSG and their `helicities' according to ref.~\cite{sie} are
$$\begin{array}[b]{ccccccccccc}\hline
e\du{\un{m}}{\a\a'} & \j\du{\un{m}}{a\a'} &
\j_{\un{m}a\a} & A_{\un{m}ab} & \c\du{abc}{\a} & \f_{abcd} & \c\du{abcde}{\a'}
& G\ud{ab}{\a'\b'} & \tilde{\j}\du{\un{m}}{a\a'} & \tilde{R}_{\un{m}\un{n}a\a'}
& \tilde{\o}_{\un{m}\a'\b'} \cr
\hline
 +2 & +3/2 & +3/2 & +1 & +1/2 & 0 & -1/2 & -1 & -3/2& -3/2
& -2 \cr
\hline
\end{array}   $$
where we have introduced the notation $\un{m}=(\m\m')$. 

The component covariant Lagrangian is given by~\cite{sie}
$$ \eqalign{
\Lag=& \ve^{\un{m}\un{n}\un{p}\un{q}}\left[
-\frac{1}{2}\tilde{\o}\du{\un{m}}{\a'\b'}T\dud{\un{n}\un{p}}{\a}{\a'}
e_{\un{q}\a\b'}+\tilde{\j}\du{\un{m}}{a\a'}\pa_{\un{n}}(e\dud{\un{p}}{\b}{\a'}
\j_{\un{q}a\b}-\j\dud{\un{p}}{b}{\a'}A_{\un{q}ba})-\frac{1}{4}
\tilde{R}\du{\un{m}\un{n}}{\a'}T\dud{\un{p}\un{q}}{a}{\a'}\right] \cr
& +e^{-1}\left[ \frac{1}{4}G^{ab\a'\b'}F_{ab\a'\b'}+\frac{1}{6}\c^{abc\a'}
\de\ud{\a}{\a'}\c_{abc\a} -\frac{1}{2304}\ve^{abcdefgh}(\de^{a\a'}\f_{abcd})
(\de_{a\a'}\f_{efgh})\right. \cr
&\left. -\ve^{absdefgh}\left( \frac{1}{192}\f_{abcd}F\du{ef}{\a\b}F_{gh\a\b}
-\frac{1}{72}\c\du{abc}{\a}\c\du{def}{\b}F_{gh\a\b}+\frac{1}{192}
\h^{mn}\f_{abcd}\c\du{efm}{\a}\c_{ghn\a}\right)\right] \cr }\eqno(A.1)$$
where the following notation has been introduced for the component torsion:
$$ \eqalign{
-T\du{\un{m}\un{n}}{\a\a'}=&\pa_{[\un{m}}e_{\un{n}]}{}^{\a\a'}+
e_{[\un{m}}{}^{\b\a'}\o_{\un{n}]\b}{}^{\a} +\j_{[\un{m}}{}^{b\a'}
\j_{\un{n}]b}{}^{\a}~,\cr
-T\du{\un{m}\un{n}}{a\a'}=&\pa_{[\un{m}}\j_{\un{n}]}{}^{a\a'}+
\h^{ab}\j_{[\un{m}}{}^{c\a'}A_{\un{n}]bc}-\h^{ab}e_{[\un{m}}{}^{\a\a'}
\j_{\un{n}]b\a}~,\cr}\eqno(A.2)$$
the supercovariant derivatives and field strengths: 
$$ \eqalign{
\de_{\a\a'}\f_{abcd}=& D_{\a\a'}\f_{abcd}-\frac{1}{6}\j\du{\a\a'[a}{\b}
\c_{bcd]\b}+\j\du{\a\a'}{e\b'}\c_{abcde\b'}~,\cr
\de_{\a\a'}\c_{abc\b}=& D_{\a\a'}\c_{abc\b}+\j_{\a\a'd\g}(\d^{\g}_{\b}
\h^{de}\f_{eabc}+\frac{1}{2}C^{\d\g}\d^d_{[a}F_{bc]\d\b})+\j\du{\a\a'}{d\b'}
\de_{\b\b'}\f_{abcd}~,\cr
F_{\un{m}\un{n}ab}=&f_{\un{m}\un{n}ab}-\j\du{[\un{m}a}{\g}\j_{\un{n}]b\g}
+ \j\du{[\un{m}}{c\a'}(e\dud{\un{n}]}{\a}{\a'}\c_{abc\a}+\frac{1}{2}
\j\dud{\un{n}]}{d}{\a'}\f_{abcd})~,\cr
F_{ab\a\b}=& e\du{\a}{\a'\un{m}}e\du{\b\a'}{\un{n}}F_{\un{m}\un{n}ab}~,\cr
F_{ab\a'\b'}=& e\udu{\a}{\a'}{\un{m}}e\du{\a\b'}{\un{n}}F_{\un{m}\un{n}ab}~,
\cr}\eqno(A.3)$$
in terms of the standard spacetime covariant derivatives 
$D_{\a\a'}=e\ud{\un{m}}{\a\a'}D_{\un{m}}$ with the self-dual part of the 
gravitational (spin) connection $\o\du{\un{m}}{\a\b}$ as in eq.~(2.10), and 
the standard $SO(8)$ Yang-Mills field strength $f_{\un{m}\un{n}ab}(A)$.

As is clear from eq.~(A.1), all the fields of negative `helicity' appear in the
Lagrangian as Lagrange multipliers. The scalars $\f_{abcd}$ are just the 
Lagrange multipliers for themselves. The vierbein $e\du{\un{m}}{\a\a'}$ 
describes on-shell a self-dual graviton, whose anti-self-dual counterpart is 
given by the abelian gauge field $\tilde{\o}_{\un{m}\a'\b'}$. The self-dual
gravitini are described on-shell in terms of the abelian gauge fermionic 
fields $\j\du{\un{m}}{a\a'}$ and $\tilde{R}_{\un{m}\un{n}a\a'}$ since the 
other fermionic fields $\j_{\un{m}a\a}$ and $\tilde{\j}\du{\un{m}}{a\a'}$ of 
helicity $\pm 3/2$ merely represent the gauge auxiliary degrees of freedom.
The equations of motion for the self-dual graviton and gravitini are just the
vanishing torsion constraints in eq.~(A.2), as they should. The vector field 
$A_{\un{m}ab}$, whose field strength is self-dual on-shell, is paired with 
its anti-self-dual tensor counterpart $G\ud{ab}{\a'\b'}$, and similarly for 
the spinor fields $\c\du{abc}{\a}$ and $\c\du{abcde}{\a'}$. 

The Lagrangian (A.1) can be considered as the $N=8$ supersymmetric covariant
generalization of the self-dual gravity in terms of Ashtekar 
variables~\cite{asht}. The corresponding $N=8$ action is invariant under the 
following {\it local} symmetries:
\begin{itemize}
\item general coordinate diffeomorphisms in the ultra-hyperbolic spacetime,
\item the $SL(2)$ local Lorentz transformations,
\item the $SO(8)$ gauge transformations,
\item $N=8$ on-shell supersymmetry, 
\item the extra abelian gauge symmetries whose gauge fields transform as \\
$\d\tilde{\o}_{\un{m}\a'\b'}=\pa_{\un{m}}\l_{\a'\b'}$, 
$\d\tilde{\j}\du{\un{m}}{a\a'}=\pa_{\un{m}}\l^{a\a'}$ and
$\d\tilde{R}_{\un{m}\un{n}a\a'}=\pa_{[\un{m}}\l_{\un{n}]a\a'}$.
\end{itemize}

The $N=8$ SDSG component action also has the {\it global} $SL(2,{\bf R})'$ 
Lorentz symmetry, and it is invariant under constant shifts of scalars 
$\f_{abcd}$ (the Peccei-Quinn-type symmetry) to be accompanied by 
corresponding transformations of some other fields too~\cite{sie}.

\end{document}